\documentclass[11pt,a4paper]{article}
\usepackage[latin1]{inputenc}
\usepackage{times}
\usepackage{epsfig}
\usepackage{amsfonts}
\usepackage{float}
\usepackage{amsmath}
\usepackage{latexsym}
\usepackage{color}
\usepackage{rotating, rotfloat}

\relax 
\allowdisplaybreaks[3]

\relax

\newcommand{\renu}{\mathbb{R}}
\newcommand{\indic}{\mathbb{I}}
\newcommand{\natnu}{\mathbb{N}}

\newcommand{\liff}{\Longleftrightarrow}

\newcommand{\nnr}{\nonumber}

\newcommand{\sep}{\quad}

\newcommand{\qed}{\mbox{ } \hfill $\Box$\\ }
\newcommand{\mc}{\multicolumn}

\newcommand{\bay}{\begin{array}}
\newcommand{\eay}{\end{array}}

\newcommand{\bqa}{\begin{eqnarray*}}
\newcommand{\eqa}{\end{eqnarray*}}

\newcommand{\bqan}{\begin{eqnarray}}
\newcommand{\eqan}{\end{eqnarray}}

\newcommand{\bit}{\begin{itemize}}
\newcommand{\eit}{\end{itemize}}

\newcommand{\ben}{\begin{enumerate}}
\newcommand{\een}{\end{enumerate}}

\newcommand{\beq}{\begin{equation}}
\newcommand{\eeq}{\end{equation}}

\newcommand{\bdes}{\begin{description}}
\newcommand{\edes}{\end{description}}

\newcommand{\btb}{\begin{tabular}}
\newcommand{\etb}{\end{tabular}}

\newcommand{\bcen}{\begin{center}}
\newcommand{\ecen}{\end{center}}

\newcommand{\bmp}{\begin{minipage}}
\newcommand{\emp}{\end{minipage}}

\newcommand{\ig}{i=1, \ldots,}
\newcommand{\jg}{j=1, \ldots,}
\newcommand{\kg}{k=1, \ldots,}
\newcommand{\rg}{r=1, \ldots,}

\newcommand{\appeq}{\mbox{ \d{$\stackrel{\textstyle .}{=}$} }}

\newcommand{\sumi}{\sum_{i=1}^}
\newcommand{\sumj}{\sum_{j=1}^}
\newcommand{\sumk}{\sum_{k=1}^}
\newcommand{\sumell}{\sum_{\ell=1}^}

\newcommand{\kp}{\otimes}

\newcommand{\cov}{\operatorname{{\it Cov}}}
\newcommand{\Cov}{\operatorname{{\it Cov}}}

\newcommand{\Var}{\operatorname{{\it Var}}}

\newcommand{\tr}{\operatorname{tr}}

\newcommand{\hc}{{\widehat c}}

\newcommand{\hF}{{\widehat F}}

\newcommand{\hG}{{\widehat G}}

\newcommand{\hp}{{\widehat p}}

\newcommand{\hs}{{\widehat s}}

\newcommand{\hv}{{\widehat v}}

\newcommand{\hw}{{\widehat w}}

\newcommand{\hsigma}{{\widehat \sigma}}

\newcommand{\htau}{{\widehat \tau}}

\newcommand{\olp}{\overline{p}}

\newcommand{\olw}{\overline{w}}

\newcommand{\olF}{\overline{F}}

\newcommand{\olR}{\overline{R}}

\newcommand{\vk}{\boldsymbol{k}}

\newcommand{\vn}{\boldsymbol{n}}

\newcommand{\vp}{\boldsymbol{p}}

\newcommand{\vt}{\boldsymbol{t}}

\newcommand{\vw}{\boldsymbol{w}}

\newcommand{\vC}{\boldsymbol{C}}

\newcommand{\vE}{\boldsymbol{E}}
\newcommand{\vF}{\boldsymbol{F}}

\newcommand{\vI}{\boldsymbol{I}}
\newcommand{\vJ}{\boldsymbol{J}}

\newcommand{\vM}{\boldsymbol{M}}

\newcommand{\vP}{\boldsymbol{P}}

\newcommand{\vS}{\boldsymbol{S}}
\newcommand{\vT}{\boldsymbol{T}}

\newcommand{\vV}{\boldsymbol{V}}
\newcommand{\vW}{\boldsymbol{W}}
\newcommand{\vX}{\boldsymbol{X}}

\newcommand{\vZ}{\boldsymbol{Z}}

\newcommand{\vmu}{\boldsymbol{\mu}}

\newcommand{\vsigma}{\boldsymbol{\sigma}}
\newcommand{\vSigma}{\boldsymbol{\Sigma}}

\newcommand{\veins}{{\bf 1}}
\newcommand{\vnull}{{\bf 0}}

\newcommand{\volR}{\boldsymbol{\overline{R}}}

\newcommand{\vwhp}{\boldsymbol{\widehat{p}}}

\newcommand{\vwhw}{\boldsymbol{\widehat{w}}}

\newcommand{\vwhF}{\boldsymbol{\widehat{F}}}

\newcommand{\vwhM}{\boldsymbol{\widehat{M}}}

\newcommand{\vwhS}{\boldsymbol{\widehat{S}}}

\newcommand{\vwhV}{\boldsymbol{\widehat{V}}}
\newcommand{\vwhW}{\boldsymbol{\widehat{W}}}

\newcommand{\vwhalpha}{\boldsymbol{\widehat{\alpha}}}

\newcommand{\whc}{\widehat{c}}

\newcommand{\whf}{\widehat{f}}

\newcommand{\whp}{\widehat{p}}

\newcommand{\whv}{\widehat{v}}
\newcommand{\whw}{\widehat{w}}

\newcommand{\whQ}{\widehat{Q}}

\newcommand{\whtau}{\widehat{\tau}}

\newcommand{\wtf}{\widetilde{f}}

\newcommand{\wtep}{\widetilde{\epsilon}}

\newtheorem{definition}{{\sc Definition}\sc}[section]
\newcommand{\bdefi}{\begin{definition}}
\newcommand{\edefi}{\end{definition}}

\newtheorem{appropr}[definition]{{\sc Approximation Procedure}\sc}
\newcommand{\bappr}{\begin{appropr}}
\newcommand{\eappr}{\end{appropr}}

\newtheorem{bedi}[definition]{{\sc Condition}\sc}
\newcommand{\bbd}{\begin{bedi}}
\newcommand{\ebd}{\end{bedi}}

\newtheorem{bedin}[definition]{{\sc Conditions}\sc}
\newcommand{\bbdn}{\begin{bedin}}
\newcommand{\ebdn}{\end{bedin}}

\newtheorem{corollary}[definition]{{\sc Corollary}\sc}
\newcommand{\bco}{\begin{corollary}}
\newcommand{\eco}{\end{corollary}}

\newtheorem{lemma}[definition]{{\sc Lemma}\sc}
\newcommand{\blem}{\begin{lemma}}
\newcommand{\elem}{\end{lemma}}

\newtheorem{proposition}[definition]{{\sc Proposition}\sc}
\newcommand{\bpro}{\begin{proposition}}
\newcommand{\epro}{\end{proposition}}

\newtheorem{satz}[definition]{{\sc Theorem}\sc}
\newcommand{\bsa}{\begin{satz}}
\newcommand{\esa}{\end{satz}}

\newtheorem{assumption}[definition]{{\sc Assumption}\sc}
\newcommand{\bas}{\begin{assumption}}
\newcommand{\eas}{\end{assumption}}

\newtheorem{assumptions}[definition]{{\sc Assumptions}\sc}
\newcommand{\bass}{\begin{assumptions}}
\newcommand{\eass}{\end{assumptions}}

\newtheorem{abb}{{\sc Figure}\sc}[section]
\newcommand{\babb}{\begin{abb}}
\newcommand{\eabb}{\end{abb}}

\newenvironment{remark}{\begin{rmk}\sl}{\end{rmk}}
\newtheorem{rmk}{{\sc Remark}\sc}[section]
\newcommand{\brem}{\begin{remark}}
\newcommand{\erem}{\end{remark}}

\newenvironment{example}{\begin{exmp}\rm}{\end{exmp}}
\newtheorem{exmp}{{\sc Example}\sc}[section]
\newcommand{\bbsp}{\begin{example}}
\newcommand{\ebsp}{\end{example}}
\newcommand{\bexa}{\begin{example}}
\newcommand{\eexa}{\end{example}}

\newtheorem{model}{{\sc Model}\sc}[section]
\newcommand{\bmdl}{\begin{model}}
\newcommand{\emdl}{\end{model}}

\newtheorem{scheme}{{\sc Scheme}\sc}[section]
\newcommand{\bscm}{\begin{scheme}}
\newcommand{\escm}{\end{scheme}}

\newenvironment{tabelle}{\begin{tabl}\sl}{\end{tabl}}
\newtheorem{tabl}{{\sc Table}\sc}[section]
\newcommand{\btab}{\begin{tabelle}}
\newcommand{\etab}{\end{tabelle}}

\newenvironment{exercise}{\begin{exc}\sl}{\end{exc}}
\newtheorem{exc}{Exercise}[section]
\newcommand{\bexe}{\begin{exercise}}
\newcommand{\eexe}{\end{exercise}}

\newcommand{\cm}{covariance matrix}
\newcommand{\cms}{covariance matrices}

\newcommand{\df}{distribution function}
\newcommand{\db}{distribution}
\newcommand{\dfs}{distribution functions}
\newcommand{\dbs}{dis\-tri\-bu\-tions}

\newcommand{\rvs}{random variables}

\newcommand{\yp}{hypothesis}
\newcommand{\yps}{hypotheses}
\newcommand{\np}{non\-pa\-ra\-me\-tric}

\newcommand{\obs}{observation}
\newcommand{\obss}{observations}

\newcommand{\asy}{asymptotic}
\newcommand{\ind}{independent}

\begin{document}

\begin{center}
{\Large \bf Rank-Based Procedures in Factorial Designs: Hypotheses about Nonparametric Treatment Effects}
\end{center}
\bcen
{\large Edgar Brunner$^{1,*}$, Frank Konietschke$^{2}$, Markus Pauly$^{3}$ and Madan L. Puri$^{4}$}\\[2mm]
\ecen
{\bf Abstract.} Existing tests for factorial designs in the nonparametric case are based on hypotheses formulated in terms of distribution functions. 
Typical null hypotheses, however, are formulated in terms of some parameters or effect measures, particularly in heteroscedastic settings. 
Here this idea is extended to nonparametric models by introducing a novel nonparametric ANOVA-type-statistic based on ranks which is suitable for testing hypotheses 
formulated in meaningful nonparametric treatment effects in general factorial designs. This is achieved by a careful in-depth study of the common distribution of 
rank-based estimators for the treatment effects. Since the statistic is asymptotically not a pivotal quantity we propose three different approximation techniques, 
discuss their theoretic properties and compare them in extensive simulations
{\color{black} together with two additional Wald-type tests. An extension of the presented idea to general repeated measures designs is briefly outlined.} 
The proposed rank-based procedures maintain the pre-assigned type-I error rate quite accurately, 
also in unbalanced and heteroscedastic models.

\vfill

\noindent{\color{black} All authors are given in alphabetic order:}\\
\noindent${}^{1}$ {Department of Medical Statistics, University of G\"ottingen, Germany}\\
\noindent${}^2$ {Department of Mathematical Sciences, University of Texas at Dallas, Dallas, USA}\\
\noindent${}^3$ {Institute of Statistics, Ulm University, Germany}\\
\noindent${}^4$ {Department of Mathematics, Indiana University, Bloomington IN, USA}

\pagenumbering{Roman}
 \newpage
 \pagenumbering{arabic}
 
 \section{Introduction} \label{intro}

Factorial designs are frequently used layouts in experimental science which are
typically inferred by means of parametric procedures. However, the corresponding
assumptions such as homoscedasticity or normality are typically not met in 
practice. Moreover, if ordinal or ordered categorical data are observed the 
classical parametric models are not appropriate since these data are non-metric 
data and means are not defined. Thus, different effect measures are required by 
which treatment effects can be described or hypotheses may be formulated. These 
nonparametric effect measures should be appropriate for metric as well as 
non-metric data, e.g., ordered categorical data. 

In a nonparametric two-sample design with \ind\ \obss\ $X_{ik} \sim F_i$, $\kg 
n_i; i=1,2$, Mann and Whitney (1947) introduced the quantity $w = P(X_{11} \le 
X_{21}) = \int F_1 dF_2$ as a \np\ measurement of an overlap of the two 
continuous \dbs\ $F_1$ and $F_2$. An estimator of $w$ is easily obtained by 
replacing the \dfs\ $F_i$ by their empirical counterparts $\hF_i$. This leads 
to the well-known rank estimator $\hw = \frac1{n_1}(\olR_{2 \cdot} - (n_2+1)/2)
$, where $\olR_{2 \cdot}$ is the mean of the ranks $R_{ik}$ of the \obss\ 
$X_{ik}$ among all $N=n_1+n_2$ \obss. The obviously appealing property that this 
estimator is obtained from the ranks of the \obss\ mainly contributed 
to the popularity of the test based on it, the so-called 
Wilcoxon-Mann-Whitney test. Moreover, as pointed out by Acion et al. (2006), 
the intuitive quantity $w$ has several desirable and meaningful properties as 
a reasonable effect for the description of the treatment and is widely accepted
in practice, {\color{black} see e.g., Brumback et al. (2006), Kieser et al. (2013), De Neve et 
al. (2014), Fischer et al. (2014), Fischer and Oja (2015) and Vermeulen et al. 
(2015). In addition, 
this effect is used for assessing the accuracy of diagnostic tests in medicine
since it is equal to the area under the receiver operating characteristic 
(ROC)-curve, see Bamber (1975) and in factorial diagnostic designs see, e.g., 
Kaufmann et al. (2005), Lange (2008), Brunner and Zapf (2013), and Zapf et al. 
(2016). 

A generalization of the relative effect $w$ to more than two \dbs\ or to 
factorial designs is not obvious and entails some difficulties. A generalization  
based on the pairwise effects $w_{\ell i}= P(X_{\ell 1} \le X_{i1}), \ell \neq 
\ig d$ has been considered by Rust and Fligner (1984) for the special case of 
the several sample design assuming continuous \dfs. Using these pairwise 
relative effects, however, can lead to paradox results since the pairwise 
effects are not transitive. For details see e.g. Gardner (1970), Brown and 
Hettmansperger (2002) or Thangavelu and Brunner (2007) and the references cited 
therein.

The problem of the non-transitivity of the pairwise effects can be circumvented 
by comparing the \dfs\ $F_i$ with the same reference \db. For several samples 
with independent observations $X_{ik} \sim F_i$, $\ig d$; $\kg n_i$, $N = 
\sumi d n_i$, Kruskal (1952) and Kruskal and Wallis (1952) used the pooled \df\ 
$H=\frac1N \sumi d n_i F_i$ as a reference \db\ and suggested the relative 
effect $r_i = \int H dF_i$ as a nonparametric effect measure. Since $H$ is the 
mean of the \dfs\ $F_i$ weighted by the relative sample sizes $n_i/N$, the 
quantities $r_i$ depend 
on the sample sizes and can therefore not be regarded as model constants by 
which \yps\ may be formulated. For this reason a different nonparametric effect 
measure $p_i = \int G dF_i$, for the case of $d$ \dfs\ $F_1, \ldots, F_d$ had 
been suggested by Brunner and Puri (2001). Here, $G = \frac{1}{d} \sum_{i=1}^d 
F_i$, denotes the unweighted mean of the \dfs\ $F_1, \ldots, F_d$. This effect 
size measure has been studied in more detail by Domhof (2001) and further by 
Gao and Alvo (2005, 2008), and Gao et al. (2008).

To demonstrate the meaning of the dependency of $r_i$ on sample sizes, consider 
the following example. Let $F_i$, $i=1,2,3$ denote normal \dbs\ $N(\mu_i,1)$ 
with expectations $\mu_1=1, \mu_2=0$, and $\mu_3=-1$ and variances 
$\sigma_i^2 \equiv \sigma^2 = 1$. Let further denote $n_{{\color{black}1}}=20, n_{{\color{black}2}}=10$, 
and $n_{{\color{black}3}}=5$, the first setting of sample sizes and $n_{{\color{black}1}}=5, n_{{\color{black}2}}=10$, 
and $n_{{\color{black}3}}=20$ the second setting where $N=35$ is the total sample size in 
both cases. Finally let $H=\frac1N \sumi d n_i F_i$ denote the weighted mean of 
the \dfs\ and $G = \frac1d \sum_{i=1}^d F_i$ the unweighted mean. The (weighted)
relative effects $r_i = \int H dF_i$ and the (unweighted) relative effects $p_i 
= \int G dF_i$ displayed in Table~\ref{relef} for the two settings of sample 
sizes are quite different. Obviously, it is not reasonable to regard the 
'effects' $r_i$ as fixed model effects by which hypotheses could be formulated 
or for which confidence intervals could be constructed. The unweighted effects 
$p_i$, however, remain unchanged by the different settings of sample sizes. Thus
these unweighted effects will be used for the formulation of hypotheses.

\begin{table}[h]
{\color{black}
\caption{Weighted relative effects $r_i = \int H dF_i$ (left) and unweighted 
relative effects $p_i = \int G dF_i$ (right) for the two settings of sample sizes 
and for the normal \dbs\ $F_1 = N(1,1)$, $F_2 = N(0,1)$, and $F_3 = N(-1,1)$.}
\label{relef}
\bcen
\btb{|l|c|c|c|c|c|c|} \hline
& \mc{3}{|c|}{Weighted Relative Effects} & 
\mc{3}{|c|}{Unweighted Relative Effects} \\ \cline{2-7}
\mc{1}{|c|}{Sample Sizes} & $r_1$ & $r_2$ & $r_3$ & $p_1$ & $p_2$ & $p_3$ \\ 
\hline
Setting 1 \sep 20, 10, 5 & 0.635 & 0.388 & 0.185 & 0.727 & 0.5 & 0.273 \\ \hline
Setting 2 \sep 5, 10, 20 & 0.815 & 0.612 & 0.365 & 0.727 & 0.5 & 0.273 \\ \hline
\etb
\ecen
}
\end{table}

The hypothesis of no treatment effect $H_0^w: w = \frac12$ in the case of two 
samples is extended to the several sample design as $H_0^p: \{ \vP_d\ \vp = 
\vnull \} = \{p_1=\dots=p_d\}$, where $\vp = (p_1, \ldots, p_d)'$ denotes the vector of the 
(unweighted) relative effects $p_i = \int G dF_i$ and $\vP_d = \vI_d - \frac1d 
\vJ_d$ denotes the centering matrix and $\vI_d$ and $\vJ_d$ the $d$-dimensional identity matrix and matrix of $1$'s, respectively. The extension to factorial designs is 
obvious. Replacing the centering matrix $\vP_d$ by an appropriate contrast 
matrix $\vC$, the hypothesis is then formulated as $H_0^p(\vC): \vC \vp = 
\vnull$. In the same way, the stronger hypothesis of no treatment effect 
$H_0^F: F_1 = F_2$ can be extended from the case of two samples to factorial 
designs by using an appropriate contrast matrix $\vC$ and stating the 
hypothesis as $H_0^F (\vC): \vC \vF = \vnull$, where $\vF = (F_1, \ldots, 
F_d)'$ denotes the vector of \dfs\ (Akritas and Arnold, 1994). These 
hypotheses, however, are more restrictive than the hypotheses formulated by the 
relative effects $p_i = \int G dF_i$ since $\vC \vF = \vnull$ implies $\vC \vp =\int 
G d(\vC \vF) = \vnull$ but not vice versa. Note that in both cases the hypotheses are based on 
fixed model quantities which do not depend an sample sizes.

The advantage of the procedures based on $H_0^F (\vC): \vC \vF = \vnull$ is 
that the covariance matrix of the contrasts $\vC \volR_\cdot$ of the vector of rank means 
$\volR_\cdot = (\olR_{1 \cdot}, \ldots, \olR_{d \cdot})'$ has a quite simple 
form under this hypothesis (Akritas et al, 1997; Akritas and
Brunner, 1997). Moreover, It can be consistently estimated from the ranks. The 
clear disadvantage is that these procedures are only designed for testing and 
that there are no fixed model quantities by which easily interpretable 
treatment effects could be defined or confidence intervals could be computed to
visualized the variability of the data in the trial.

This would be different for procedures based on the (unweighted) relative 
treatment effects $p_i$, which are appropriate to describe nonparametric 
treatment effects for which confidence intervals could be derived. On the other
hand, the covariance matrix of the contrasts $\vC \volR_\cdot$ of the rank 
means has a quite involved structure under the hypothesis $H_0^p(\vC): \vC \vp 
= \vnull$ (for details see Puri, 1964, who derived the general \cm\ of the 
vector of rank means $\volR_\cdot$). This fact seems to be one of the reasons 
why general rank tests in factorial designs have mainly been developed for 
testing hypotheses based on the \dfs, i.e. $H_0^F(\vC): \vC \vF = \vnull$. 
It is our intention to close this gap. The computation of the quite involved 
covariance matrix is based on a similar matrix technique as used in Konietschke et al. (2012) which is generalized here to factorial designs. 
Moreover, we want to provide procedures for testing hypotheses about easily 
interpretable nonparametric treatment effects in those cases where it is 
obviously not reasonable to formulate hypotheses based on contrasts of \dfs. 
This may be regarded as a generalization of the Behrens-Fisher problem to 
nonparametric factorial designs. A real data set is provided in the following 
example.

The immune system stimulating effect in a stress situation of a drug compared 
to a placebo was investigated in an animal experiment involving 40 Wistar rats 
who were randomly partitioned in two groups. One group of 20 animals received 
normal food while the other half received reduced food to generate a stress for 
these animals. Within each group 10 animals were randomly assigned to a drug 
added to the food while the other 10 rats in each group received a placebo. The 
immune system stimulating effect was measured by the number of leucocytes 
$[10^6/ml]$ in the peritoneal liquid obtained by a stimulation prior to the 
section of the abdominal membrane. This experiment was performed as a $2\times 
2$-design to answer the question whether the immune response in a stress 
situation was the same as in a normal situation. The data are displayed as box 
plots in Figure\ref{boxplots}. Obviously, it is not reasonable to formulate the hypothesis of no interaction between the stress situation and the treatment by means of the distributions functions since the variances as well as 
the shapes of the distributions are different for the two stress situations. 
Here it seems to be more appropriate to formulate the hypotheses in terms of the relative effects $p_i$. 
The {\color{black} complete data set of this example is displayed in the supplemenentary material while the} 
analysis of this experiment by the proposed rank procedure is provided in Section~\ref{exa}.

\begin{figure}
\bcen
\includegraphics[height=40ex,width=50ex]{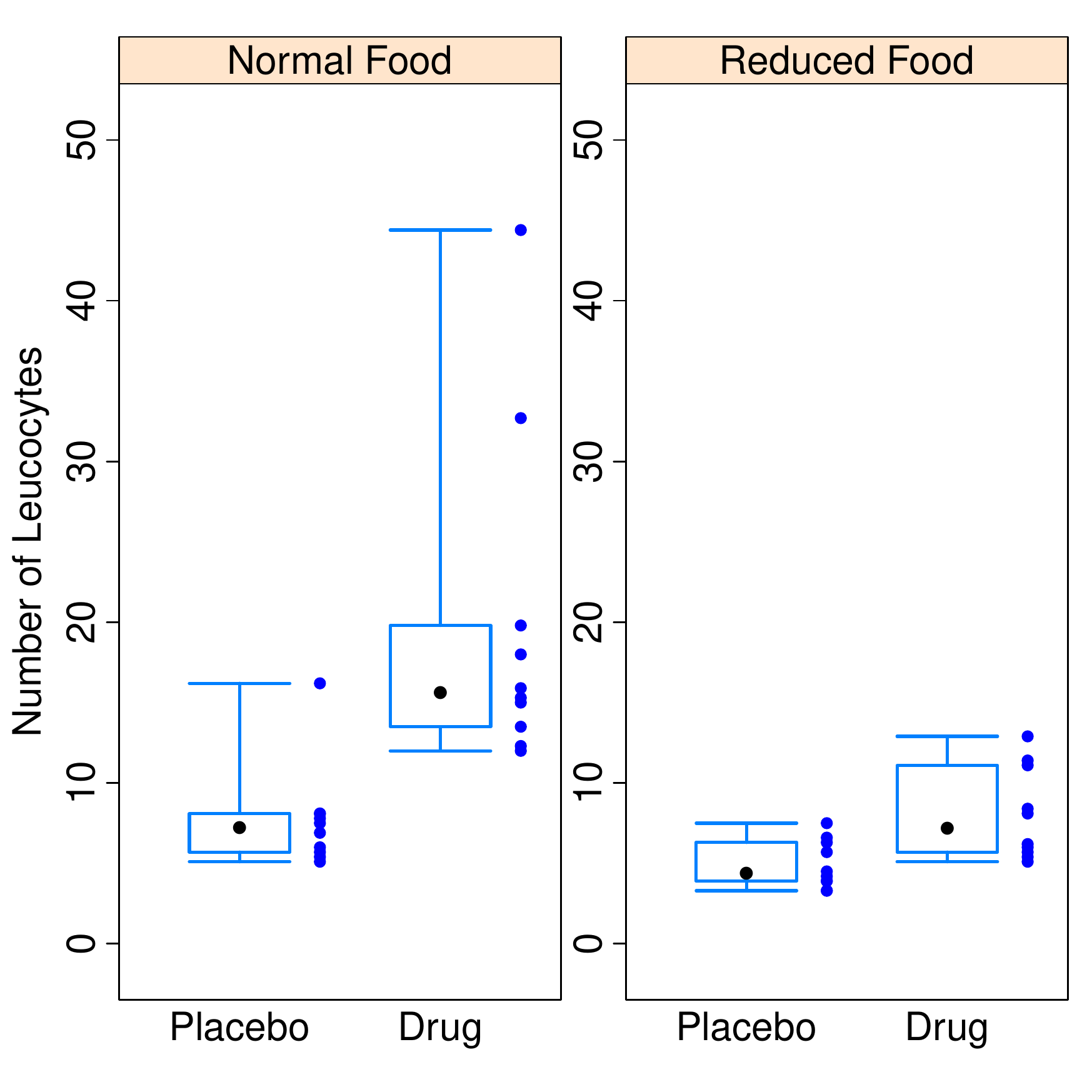}
\ecen
\caption{Box plots of the number of leucocytes $[10^6/ml]$ in the peritoneal 
liquid of Wistar rats in a stress situation (reduced food) compared to a normal 
situation (normal food) treated by a drug or by a placebo.} \label{boxplots}
\end{figure}

Recently, {\color{black} Fan and Zhang (2014, 2015) have proposed a GEE approach for rank transformed data 
and} Thas et al. (2012), and De Neve and Thas (2015) have introduced a similar concept of the so-called probabilistic 
index models (PIM). It allows for flexible rank-based 
modeling for various designs. These models, however, are based on weighted 
effects, where sample sizes are involved and the related inference procedures 
are mainly developed for null hypotheses formulated in terms of the \dfs\ {\color{black} (with the exception of De Neve and Thas, 2015, who considered 
null hypotheses in terms of probalistic indices for special models such as the unpaired two-sample design)}.

Moreover, all above mentioned procedures are based on asymptotic considerations 
while in general, approximations for small samples have only been investigated 
for $H_0^F(\vC)$ in Brunner et al. (1997, 1999) and Brunner 
and Langer (2000). Thus, it is the intention of the present paper to derive
 inference procedures for $H_0^p(\vC)$ which are asymptotically valid and at the same time
possess good small sample properties. 
This appears particularly necessary since most of the biological or medical
 experiments are performed as laboratory or animal 
experiments or as trials in clinical epidemiology involving only a small or 
moderate number of animals or patients or with biological material where only a 
few {\color{black}replicates} are available.

The paper is organized as follows. In Section~\ref{mod} we introduce the
 underlying model, effects and corresponding rank estimators. After studying 
their asymptotic distribution in Section~\ref{asy}, we propose statistics for
 testing factorial hypotheses about the nonparametric treatment effects and
 investigate their asymptotic properties in Section~\ref{stat}. Section~\ref{erm} generalizes the current approach to repeated measures designs. Applications to
 specific layouts  are given in Section~\ref{spd}. In Section~\ref{sim} extensive 
simulations are conducted and different approximation techniques are compared 
with the classical Kruskal-Wallis test. The results obtained by 
the rank-based procedures suggested in this paper are discussed in the last section. All proofs and technical details are given in the Appendix.

\section{Notations and Statistical Model}  \label{mod}

To be as general as possible we assume a nonparametric model with independent \rvs\ 
\bqan \label{model}
X_{ik} \sim F_i, \quad \ig d; \ \kg n_i.
\eqan
In order to allow for ties and all types of discrete data, such as count data, ordered categorical data and even 
dichotomous data in a unified form, we use the so-called normalized 
version of the \df\ $F_i = \frac12 (F_i^+ + F_i^-)$ which is the mean of 
its right- and left-continuous version (Ruymgaart, 1980; Akritas et al., 1997; Munzel, 1999). To describe treatment effects in this set-up, 
we will use the relative effects {\color{black} between the distributions of group $i$ and $j$}
\bqan \label{wli}
w_{j i} &=& P(X_{j 1} < X_{i1}) + \tfrac12 P(X_{j 1} = X_{i1}) 
\ = \ \int F_j d F_i, \sep j, \ig d
\eqan
and define a relative treatment effect of \db\ $i$ with respect to all 
\dbs\ $\jg d \ $ by 
\bqan \label{releff}
p_i &=& \int G dF_i \ = \ \olw_{\cdot i},
\eqan
where $G=\frac1d \sum^d_{\color{black} i= 1} F_{\color{black}i}$ is the unweighted mean of the \dfs\ 
$F_i$, $\ig d$. 
{\color{black} In particular, 
$p_i = P(Z < X_{i1}) + \frac12 P(Z= X_{i1})$ for a random variable $Z\sim G$ being independent of $X_{i1}$.
}{\color{black} 
 Thus, an effect $p_i$ smaller than $1/2$ means that the observations from the distribution $F_{i}$ tend to smaller values than those from the mean distribution $G$. 
}Note that by definition, $\frac{1}{2d}\leq p_i\leq 1-\frac{1}{2d}$.

The use of $p_i$ allows for a transitive effect ordering which would 
in general not be the case with $w_{\ell i}$, see the discussion in Brown and Hettmansperger (2002).
Moreover, these relative effects do not have the drawback of 
depending on sample sizes $n_1, \ldots, n_d$ unlike the quantities $r_i = 
\int H dF_i$ defined by Kruskal and Wallis (1952), where $H=\frac1N 
\sumell d n_\ell F_\ell$ denotes the weighted mean of the \dbs. 
This 
advantage enables the formulation of \np\ \yps\ in terms of these 
relative effects in a general set-up. To this end let $\vp = (p_1, \ldots 
p_d)'$ denote the vector of these relative effects and let $\vC$ denote 
an appropriate contrast matrix to formulate any linear \yp\  
\bqan\label{H0p}
H_0^p: \vC 
\vp = \vnull
\eqan
about the relative treatment effects defined in \eqref{releff}.
We note that factorial designs are covered by this approach by 
introducing an appropriate structure for the index $i$ by splitting it in 
sub-indices $i_1, i_2, \ldots$ according to the number of factors 
considered in the design. Thus, this formulation includes nested designs as 
well as cross-classifications of different orders, see, e.g., Section~2.1 in Brunner and Puri (2001) in connection with $H_0^F$ or 
Section~4 in Pauly et al. (2015a) 
for such designs in a semiparametric framework that can be directly translated to our model \eqref{model}.

Let $\hF_i(x) = \frac1{n_i} \sumk {n_i} c(x-X_{ik})$ denote the empirical 
\df\ of $F_i(x)$, where $c(u)$ denotes the normalized version of the count 
function, i.e. $c(u)=0, \frac12,1$ according as $u<, =,$ or $>0$. 
Replacing the \dfs\ $F_i(x)$, $\ig d$, by their empirical counterparts 
$\hF_i(x)$, estimators of the the relative effects $p_i$ are obtained from 
linear combinations of all pairwise rankings, i.e.
\bqan \label{releffest}
 \hp_i &=& \int \hG d \hF_i \ = \ \frac1d \sumell d \int \hF_\ell d \hF_i 
 \ = \ \frac1d \sumell d \hw_{\ell i} \ ,
\eqan
where 
\bqan \label{rankrep} 
\hw_{\ell i} &=& \frac1{n_\ell} \left( \olR_{i \cdot}^{(\ell i)} - 
\frac{n_i+1}2 \right), \sep \olR_{i \cdot}^{(\ell i)} \ = \ \frac1{n_i}
\sumk {n_i} R_{ik}^{(\ell i)}
\eqan
and where $R_{ik}^{(\ell i)}$ denotes the (mid-)rank of $X_{ik}$ among all 
$n_\ell + n_i$ \obs\ within the two samples $X_{\ell 1}, \ldots, 
X_{\ell n_\ell}, X_{i1}, \ldots X_{i n_i}$. Let $\vwhp = (\hp_1, \ldots, 
\hp_d)'$ denote the vector of the $d$ estimated relative effects. Note 
that in case of equal sample sizes $n_1= \cdots = n_d=n$, the quantities 
$\hp_i$ reduce to $\frac1N (\olR_{i \cdot} - \frac12)$, where $R_{ik}$ is the rank of $X_{ik}$ in the combined sample and 
$\olR_{i \cdot} = \frac1{n_i} \sumk {n_i} R_{ik}$ are the the rank means. It is well-known that also in the case of ties the estimates $\hw_{\ell i}$ are 
unbiased and $L_2$-consistent estimators of $w_{\ell i}$, $\ell, \ig d$, see,
e.g., Brunner and Puri (2001). The same properties also hold for the estimators 
$\hp_i$ since they are linear combinations of the $\whw_{\ell i}$. 

The representation of the asymptotic \cm\ of $\sqrt{N} \vwhp$, however, is quite
involved (Puri, 1964) and the derivation of consistent estimators of the 
variances and the covariances requires tedious computations. Therefore, it is 
one of the aims of the present paper to provide a simple technique for the 
representation and estimation of theses quantities. To this end, we will use the
following vector and matrix notation. Let $\vF = (F_1, \ldots, F_d)'$ denote the
vector of the \dfs\ and let $$\vw_i = (w_{1 i}, \ldots, w_{d i})' = \int \vF 
dF_i $$ and $\vw =(\vw_1',\vw_2', \dots, \vw_d')'$ denote the $d^2$-vector of 
the relative effects $w_{\ell i}$ in (\ref{wli}). The estimators $\vwhw$ and 
$\vwhw_i = \int \vwhF d \hF_i$ are defined accordingly. Finally, let $\vE_d = 
\vI_d \kp \frac1d \veins_d'$ where $\vI_d$ denotes the $d$-dimensional unit 
matrix, $\veins_d=(1, \ldots,1)'$ the $d \times 1$ vector of $1$s and $\kp$ 
denotes the Kronecker product of matrices. Then the vector $\vp$ of the relative
effects and its estimator $\vwhp$ can be represented as
\bqan \label{vecrep}
\vp \ = \ \vE_d \cdot \vw & \text{and} & \vwhp \ = \ \vE_d \cdot 
\vwhw,
\eqan
By (\ref{vecrep}), the \asy\ 
\cm\ $\vV$ of $\sqrt{N} (\vwhp - \vp)$ can be represented as $\vV = 
\vE_d \vS \vE_d'$, where $\vS$ denotes the \asy\ \cm\ of $\sqrt{N} 
\vwhw$. Note that $\vS = (\vS_{ij})_{i,\jg d}$ is a $d^2 \times 
d^2$ partitioned matrix the elements $\vS_{ij} \in \renu^{d \times d}$ of 
which are the \asy\ \cms\ of $\sqrt{N} (\vwhw_i', \vwhw_j')'$. 
The representation and estimation of the matrices 
$\vS_{ij}$ are discussed in the next section.

\section{Asymptotic Results} \label{asy}

Here we derive the \asy\ \db\ of $\vt_N = \sqrt{N} (\vwhp - \vp)$ under the following framework
\bqan\label{ass: asymp}
\min_{1\leq i\leq d}(n_i) \to \infty \quad \text{such that} \quad N/n_i\leq N_0<\infty \quad\text{ for all } \quad i=1,\dots,d.
\eqan
We first summarize some well-known results about the \asy\ \db\ of $\sqrt{N} (\hw_{\ell i} - w_{\ell i}) = 
\sqrt{N} \left( \int \hF_\ell d \hF_i - \int F_\ell d F_i \right)$, 
where $N = \sumi d n_i$, see, e.g., Brunner and Munzel (2000). 
To this end we re-state the \asy\ equivalence theorem for $\whw_{\ell i}$ 
to prepare the more involved \asy\ results for $\vt_N$. This theorem 
represents $t_{N}(\ell,i)=\sqrt{N} (\hw_{\ell i} - w_{\ell i})$ by sums of \ind\ 
\rvs\ $U_N(\ell,i)$ which have, \asy ally, the same \db\ as $t_{N}(\ell,i)$. Notice that we neither 
require that the \dbs\ $F_\ell$ and $F_i$ are continuous nor that the 
ratios of the sample sizes converge to constants. For convenience, we only formulate the results for $\ell=1$ and $i=2$. 
All other combinations of $(\ell,i), \ell\neq i,$ follow immediately. 

\bsa[Asymptotic Equivalence] \label{aet2}
Let $X_{jk} \sim F_j = \frac12 [F_j^+ + F_j^-]$ be \ind\ \obss, $j=1,2; 
\ \kg n_j$. Let $w_{12}$ and $\hw_{12}$ be as 
defined in (\ref{wli}) and (\ref{rankrep}), respectively. If 
$\min(n_1,n_2) \to \infty$ then $\sqrt{N} (\hw_{12} - w_{12})$ has, \asy ally, the same \db\ as 
\bqan \label{asyequiv}
U_N = U_N(1,2) &=& \sqrt{N} \left( \frac1{n_2} \sumk {n_2}[F_1(X_{2k}) - w_{12}] - 
\frac1{n_1} \sumk {n_1} [F_2(X_{1k}) - w_{21}] \right)
\eqan
{\color{black} which is a sum of (unobservable) \ind\ \rvs }.
\esa

From this we can directly deduce the \asy\ normality of $\sqrt{N} (\hw_{12} - w_{12})$ from the Central Limit Theorem.

\bsa[Asymptotic Normality] \label{asynor}
Let $\sigma_N^2 = N[\sigma_{1}^2/n_1 + \sigma_{2}^2/n_2]$ denote the variance of $U_N$, where 
$\sigma_{1}^2 = \Var(F_2(X_{11}))$ and $\sigma_{2}^2 = \Var(F_1(X_{21}))$ and assume that 
$\sigma_{1}^2, \sigma_{2}^2 >0$. Then, under the assumptions of Theorem~\ref{aet2}, 
$\sqrt{N} (\hw_{12} - w_{12}) / \sigma_N$ has, \asy ally, a standard 
normal \db\ $N(0,1)$. 
\esa

An $L_2$-consistent rank estimator $\hsigma_N^2$ is given below.

\bsa[Variance Estimator] \label{asyvar12}
Under the assumptions of Theorem~\ref{aet2} and Theorem~\ref{asynor}
an $L_2$-consistent estimator of the unknown variance $\sigma_N^2$ 
is given by $\hsigma_N^2 = N [\hsigma_{1}^2/n_1 + \hsigma_{2}^2/n_2]$, 
where 
\bqan \label{varest2}
\hsigma_{j}^2 &=& \frac1{(n_1+n_2-n_j)^2(n_j-1)} \sumk {n_j} \left( 
R_{jk}^{(12)} - R_{jk}^{(j)} - \olR_{j \cdot}^{(12)} + \tfrac{n_j+1}2 
\right)^2, \; j=1,2
\eqan
is the empirical variance of $\frac1{n_1+n_2-n_j} \left( R_{jk}^{(12)} - 
R_{jk}^{(j)}\right)$. Here, $R_{jk}^{(12)}$ denotes the (mid-)rank of 
$X_{jk}$ among all $n_1+n_2$ \obss\ in the pooled samples 1 and 2, while 
$R_{jk}^{(j)}$ denotes the (mid-)rank of $X_{jk}$ among all $n_j$ 
\obss\ within sample $j$, for $j=1,2$.
\esa

{\color{black} 
For a proof of the results stated in Theorems~\ref{aet2} and ~\ref{asyvar12} we refer to Brunner and Munzel (2000).
} From these theorems it follows immediately that $\sqrt{N} (\hw_{12} - 
w_{12}) / \hsigma_N$ has, \asy ally, a standard normal \db\ $N(0,1)$.

To derive the \asy\ \cm\ of $\vt_N=\sqrt{N}(\vwhp - \vp)$, we use the 
representation of $\vwhp$ and $\vp$ in (\ref{vecrep}) and obtain
\bqan \label{asyvecrep}
\sqrt{N}(\vwhp - \vp) &=& \vE_d \cdot \left( \sqrt{N} (\vwhw - 
\vw) \right).
\eqan
From Theorem~\ref{asyvar12} we obtain the following asymptotic equivalence for the components $\hw_{\ell i} - w_{\ell i}$ of $\vwhw-\vw$,
\bqa
 \sqrt{N} (\hw_{\ell i} - w_{\ell i}) & \appeq & \sqrt{N} \left[ 
 \frac1{n_i} \sumk {n_i}[F_\ell(X_{ik}) - w_{\ell i}] - \frac1{n_\ell} 
 \sumk {n_\ell} [F_i(X_{\ell k}) - w_{i \ell}] \right] \\
 & =: & \sqrt{N} Z_{\ell i},
\eqa
where $N=\sumi d n_i$ and the sign $\appeq$ means that the two 
sequences of \rvs\ on the left and right side are \asy ally equivalent. 
This means in particular that they have the same \asy\ \cm. 

We collect the unobservable \rvs\ $Z_{\ell i}$ in the vectors $\vZ_i = (Z_{1i}, 
\ldots, Z_{di})',$ $\ig d$ and let $\vZ = (\vZ_1', \ldots, \vZ_d')'$. Thus, by definition,
$E(\vZ)=\vnull$ and $Z_{ii}=0$. Then the asymptotic equivalence $\sqrt{N}(\vwhp - \vp) \appeq 
\vE_d \cdot \sqrt{N} \vZ$  follows from (\ref{asyvecrep}).

Next, we derive the \asy\ \cm\ $\vS$ of $\sqrt{N} \vZ$. According 
to partitioning $\vZ$ in $\vZ_1, \ldots, \vZ_d$, we also partition 
$\vS$ as $\vS=(\vS_{ii'})_{i,i'=1, \ldots, d}$, where $\vS_{ii} = 
\Cov(\sqrt{N} \vZ_i)$ and $\vS_{ii'} = \Cov(\sqrt{N} \vZ_i, \sqrt{N} 
\vZ_{i'})$, $i \neq i'$.

Let $s_i(\ell, \ell')$ denote the $(\ell, \ell')$-element of 
$\vS_{ii}$ and $s_{ii'}(\ell, \ell')$  the $(\ell, \ell')$-element of 
$\vS_{ii'}$ for $i \neq i'$. Then, by independence of $X_{ik}$ and 
$X_{i'k'}$ if $(i,k) \neq (i',k')$, we obtain the elements of $\vS_{ii}$
\bqan \label{sill}
s_i(\ell, \ell') &=& \left\{\bay{ll} \tau_i(\ell, \ell) + \tau_\ell(i,i)
& \ell = \ell', \ \ell \neq i \\
\tau_i(\ell, \ell'), &  \ell \neq \ell', \ i \neq \ell, \ i \neq \ell' \\
0, & \text{if otherwise} 
 \eay
\right.  \nnr \\[-6mm]
\eqan
and of $\vS_{ii'}$ for $i \neq i'$
\bqan \label{siill}
s_{ii'}(\ell, \ell') &=& \left\{\bay{ll} 
- \tau_i(\ell, i'), & \ell \neq \ell', \ell \neq i', \ell' = i \\
- \tau_i(i',i') - \tau_{i'}(i,i), & 
\ell \neq \ell', \ell = i', \ell' = i \\
- \tau_\ell(i,\ell') & \ell \neq \ell', \ell = i', \ell' \neq i \\
\tau_\ell(i,i') & \ell = \ell', \ell \neq i', \ell' \neq i \\
0, & \text{if otherwise}
\eay
\right. \nnr \\[-6mm]
\eqan
where \\[-7mm]
\bqan \label{tau}
\tau_r(s,t) &=& \frac{N}{n_r} E \left[ \left(F_s(X_{r1}) - w_{sr} \right) 
\left(F_t(X_{r1}) - w_{tr} \right) \right].
\eqan

The explicit derivation of these formulas is given in the Appendix. Finally, we obtain the asymptotic covariance matrix of $\sqrt{N} (\vwhp - \vp)$ as $\vV = (v_{ij})_{1\leq i,j\leq d},$ 
where $v_{ij}=\veins_d'\vS_{ij}\veins_d/d^2$.

The unknown quantities $\tau_r(s,t)$ in (\ref{tau}) are easily 
estimated from the pairwise ranks of the samples $s, \rg d$, which are 
obtained by replacing the \dfs\ in (\ref{tau}) with their empirical 
counterparts. Let $R_{rk}^{(sr)}$ denote the (mid-)rank of $X_{rk}$ 
among all $n_s + n_r$ \obss\ within the samples $s$ and $r$ and let 
$R_{rk}^{(r)}$ denote the (mid-)rank of $X_{rk}$ among all $n_r$ \obss\ 
within sample $r$. Finally, let $\olR_{r \cdot}^{(sr)}$ denote the mean 
of the ranks $R_{rk}^{(sr)}$ of the $n_r$ \obss\ within sample $r$. 
Then, by using basic relations between ranks and empirical \dfs\ it 
follows that
\bqan \label{fsrk}
\hF_s(X_{rk}) - \hw_{sr} &=& \frac1{n_s} \left[ \left( R_{rk}^{(sr)} - 
R_{rk}^{(r)} \right) - \left( \olR_{r \cdot}^{(sr)} - \tfrac{n_r+1}2 
\right) \right].
\eqan

Now let $D_{rk}(s) = \hF_s(X_{rk}) - \hw_{sr}$ for convenience. Then 
a rank estimator of $\tau_r(s,t)$ in (\ref{tau}) is given by
\bqan \label{tauest}
\htau_r(s,t) &=& \frac{N}{n_r(n_r-1)} \sumk {n_r} D_{rk}(s) \cdot 
D_{rk}(t).
\eqan

The $L_2$-consistency of these estimators is easily established by the 
same techniques used to prove Theorem~\ref{asyvar12} (see, e.g, Brunner 
and Munzel, 2000). The details are therefore omitted. 

Replacing the quantities $\tau_r(s,t)$ in (\ref{sill}) and (\ref{siill}) 
by $\htau_r(s,t)$, we obtain $L_2$-consistent estimators $\hs_i(\ell, 
\ell')$ and $\hs_{ii'}(\ell, \ell')$ for the covariance elements given in (\ref{sill}) and (\ref{siill}), respectively. The resulting estimator of the 
\asy\ \cm\ $\vS$ of $\sqrt{N} \vZ$ is denoted by $\vwhS_N = 
(\vwhS_{ii'})_{i,i'=1, \ldots, d}$ and finally, from (\ref{asyvecrep}), 
we obtain an estimator $\vwhV_N$ of the \asy\ \cm\ $\vV$ of the statistic 
$\vt_N = \sqrt{N} (\vwhp - \vp)$ 
\bqan \label{covestt}
\vwhV_N &=& \vE_d \cdot \vwhS_N \cdot \vE_d' = (\hv_{ij})_{1\leq i,j\leq d}, 
\eqan
where $\hv_{ij} = \veins_d'\vwhS_{ij}\veins_d/d^2$.

\section{Test Statistics} \label{stat}
The considerations from the preceding sections show that $\sqrt{N} (\vwhp - \vp)$ is asymptotically normal with mean zero and 
covariance matrix $\vV= \vE_d \cdot \vS \cdot \vE_d'$. This allows for constructing approximate test procedures for the null hypothesis $H_0^p :\vC \vp=\vnull$ stated in \eqref{H0p}. 
Let ${\color{black} \vM^+}$ denote the Moore-Penrose inverse of a matrix ${\color{black} \vM}$. For testing  $H_0^p$ the so-called Wald-type statistic {\color{black} (WTS)}
\bqan \label{WTS}
 {\color{black} W_N}(\vC) &=& N \cdot \vwhp' \vC' ( \vC \vwhV_N \vC')^+ \vC \vwhp
\eqan
may be utilized. Recall, however, that even for the more restrictive null hypothesis $H_0^F$ it is well known that ${\color{black} W_N}(\vC)$ may become extremely liberal 
unless very large sample sizes are available, see, e.g., Brunner et al. (1997) or Vallejo et al. (2010). 
In our situation this becomes even worse due to the complicated structure of the covariance matrix $\vV$ involving more unknown quantities which have to be estimated. 
{\color{black} Moreover, the matrix $\vV$ is in general singular since the sum of all effects is always 
constant $\sum_{i=1}^d \whp_i = \sum_{i=1}^d p_i  = d/2$. Thus, if $\vC$ does not have a full column rank it follows that the matrix
${\color{black}\vwhM_N^+} = ( \vC \vwhV_N \vC')^+$ does in general not converge in probability to the matrix ${\color{black} \vM^+}=(\vC \vV \vC')^+$. Therefore ${\color{black} W_N}(\vC)$ is not asymptotically 
$\chi^2_{r(\vC)}$-distributed under $H_0^p$. Assuming that $r({\color{black}\vwhM_N}) \to r({\color{black} \vM})$ in probability, an asymptotically 
valid Wald-type test for 
$H_0^p$ is given by comparing ${\color{black} W_N}(\vC)$ with the $(1-\alpha)$-quantile of a $\chi^2_{r({\color{black}\vwhM_N})}$-distribution. 
However, the above rank assumption may be hard to justify in practice. 
} 
For these reasons we do not consider this approach. Instead we use the ANOVA-type-statistic 
\bqan \label{ATS}
 {\color{black} Q_N}(\vC) \ = \ {\color{black} Q_N}(\vT) &=&  \frac{N }{\tr(\vT \vwhV_N)} \vwhp' \vT \vwhp, 
\eqan
where $\vT=\vC'(\vC\vC')^+\vC$ is the unique projection matrix on the column space of $\vC$, see e.g. Brunner et al. (1997) or Brunner and Puri (2001). Note
that $H_0^p :\vT \vp = \vnull \ \liff\ \ \vC \vp = \vnull$ since $\vC' (\vC 
\vC')^+$ is a generalized inverse of $\vC$. Furthermore, we have implicitly 
assumed in (\ref{ATS}) that $\tr(\vT \vV)\neq 0$. This assumption is quite weak
and simply means that the projection of $\vwhp$ into the hypothesis space
is non-constant (almost surely).

\bsa[Asymptotic Distribution of ${\color{black} Q_N}(\vT)$] \label{asFN}
Under the assumptions \eqref{model} and \eqref{ass: asymp} the quadratic form 
${\color{black} Q_N}(\vT)$ in (\ref{ATS}) has, asymptotically under the null $H_0^p :\vT \vp=\vnull$, the same distribution as 
\bqan\label{wtsp}
  Q(\vT) &=& \sum_{i=1}^d \frac{\lambda_i(\vT \vV)}{\tr(\vT \vV)} C_i^2,
\eqan
  where 
$C_i$ are independent standard normal random variables and $\lambda_i(\vT \vV), 1\leq i\leq d,$ denote the eigenvalues of $\vT \vV$.
\esa

Since the asymptotic distribution of ${\color{black} Q_N}(\vT)$ is non-pivotal we propose three 
different approximation procedures. The first one is based on estimating the 
unknown quantities of the limit $Q(\vT)$ in Theorem~\ref{asFN} 
above by substituting $\vV$ with $\vwhV_N$, i.e. we estimate the eigenvalues
by  $\widehat{\lambda}_i = \lambda_i(\vT \vwhV_N)$, the eigenvalues of $\vT 
\vwhV_N$, and the trace by $\tr(\vT \vwhV_N)$. Then the distribution of 
\bqan \label{ATSlambda}
 \widehat{Q}(\vT)= \sum_{i=1}^d \frac{{\lambda}_i(\vT \vwhV_N)}{\tr(\vT \vwhV_N)} C_i^2
\eqan
can be calculated, e.g., via Monte-Carlo, with arbitrary precision. Denoting the corresponding $(1-\alpha)$-quantile of the distribution of $\whQ(\vT)$ by 
$\whc(\alpha)$, we obtain an asymptotic level $\alpha$ test 
$\widehat{\varphi}_N = \indic \{ {\color{black} Q_N}(\vT) > \whc (\alpha)\}$ under the null, see
Theorem~\ref{astests} below, which may be called {\it 
ANOVA-eigen-type-$p$-test}. Here, $\indic \{ \cdot\}$ denotes the indicator
function.

The second possibility is adopted from the semiparametric mean-based case in 
Brunner et al. (1997), where a well established Box-type approximation for 
quadratic forms (see Box, 1954) is used. It is obtained by fitting the first
two moments of $\vwhp' \vT \vwhp$ with that of a scaled 
$g\chi_f^2$-distribution. This leads to the following approximation
\bqan\label{eq: ANOVA}
 {\color{black} Q_N}(\vT) &\approx&  \chi^2_{f}/f,
\eqan
where $gf=\tr(\vT\vV)$ and $f$ is estimated by
\bqan\label{eq: df2}
\widehat{f} = \frac{\tr^2(\vT \vwhV_N)}{tr(\vT\vwhV_N\vT\vwhV_N)}.
\eqan

This leads to the test $\widetilde{\varphi}_N = \indic \{\whf {\color{black} Q_N}(\vT) > 
\chi^2_{\widehat{f},1-\alpha}\}$ which may be called {\it 
ANOVA-Box-type-$p$-test}. In order to correct for a slight liberality of this
test for small sample sizes we approximate the null distribution of ${\color{black} Q_N}(\vT)$ 
by an $F(f,f_1)$-distribution. The second degree of freedom $f_1$ is chosen in 
such a way that asymptotically the approximation in \eqref{eq: ANOVA} is 
obtained. Moreover, in the two-sample case it reduces to the approximation 
given in Brunner and Munzel (2000). To this end, we suggest to estimate $f_1$ by
\bqan \label{whf1}
\whf_1 = \frac{\left[ \sum_{i=1}^d S_i^2 / (N-n_i)\right]^2}{\sum_{i=1}^d [S_i^2/(N-n_i)]^2/(n_i - 1)},
\eqan
where
\bqa
S_i^2 = \frac{1}{n_i-1}\sum_{k=1}^{n_i} \left( R_{ik} - R_{ik}^{(i)} - \olR_{i\cdot} + \frac{n_i+1}{2}\right)^2.
\eqa

Here, $R_{ik}$ 
denotes the rank of $X_{ik}$ among all $N$ observations and $R_{ik}^{(i)}$ the rank of $X_{ik}$ among all $n_i$ observations in group $i$. Note that the estimator $\whf_1$ in \eqref{whf1} fulfills the desired properties, 
i.e. $\whf_1 \to \infty$ in probability under \eqref{ass: asymp} and reduces to the Brunner and Munzel (2000) approximation in the two-sample case.

Finally, the resulting {\it ANOVA-type-$p$-test} is given by $\varphi_N = 
\indic \{{\color{black} Q_N}(\vT) > F_{1-\alpha}(\whf, \whf_1)\}$.

\bsa[Properties of the tests]\label{astests}\text{ }\ Under the assumptions 
\eqref{model} and \eqref{ass: asymp} the following statements hold: \\
(a) The test $\widehat{\varphi}_N$ is an asymptotic level-$\alpha$ test, i.e. 
under the null $H_0^p : \vC \vp=\vnull,$ we have $E(\widehat{\varphi}_N)
\rightarrow\alpha$.\\[0.5ex]
(b) All three tests considered, namely 
\ben
\item $ \widehat{\varphi}_N = \indic \{ {\color{black} Q_N}(\vT) > \hc(\alpha)\}$, 
\item $\widetilde{\varphi}_N= \indic \{\widehat{f} {\color{black} Q_N}(\vT) > 
      \chi^2_{\widehat{f},1-\alpha}\},$ 
\item ${\varphi}_N =  \indic \{{\color{black} Q_N}(\vT) > F_{1-\alpha}(\whf, \whf_1)\}$
\een
are consistent for fixed alternatives $\vC \vp \neq \vnull$.
\esa

We would like to note that it is straightforward to derive confidence intervals 
for the nonparametric effects $p_i = \int G\ dF_i$ in \eqref{releff} and 
contrasts of them using the above results. Applying the delta-method these can 
even be made range-preserving by means of logit or probit transformations (see,
e.g., Brunner and Munzel, 2013, p.~117). For example, approximate $(1-\alpha)$ confidence intervals for $p_i$ are obtained from 
\bqan\label{eq:CIg}
CI_{g,i} =\left[ g^{-1}\left( g(\whp_i)\pm \frac{z_{1-\alpha/2}}{\sqrt{N}} \sqrt{\whv_{ii}} g'(\whp_i)\right)\right],
\eqan
where $g(\cdot)$ is differentiable in $p_i$, and with $g'(p_i)\not = 0$. For instance, $g(x)=x$, or $g(x) = logit(x)$ are typical choices (Konietschke et al., 2012).

{\color{black}

\section{Extensions to General Repeated Measures Designs} \label{erm}

The previous sections dealt with general nonparametric factorial designs 
{involving} an arbitrary number of fixed crossed or nested factors. 
In this section we outline how to generalize our idea to nonparametric factorial repeated measures designs. {This extends the results for simple one-group 
layouts by Konietschke et al. (2010) to several group layouts and general split-plot designs.}
To this end, we consider independent random vectors
\bqan \label{model rm}
\vX_{ik} = (X_{i\ell k})_{\ell=1}^t = (X_{i1k},\dots,X_{itk})', \quad \ig d; \ 
\kg n_i
\eqan
representing $t\in\natnu$ repeated measurements observed on subject $k$ in 
group $i$. As above, a factorial structure on the groups (whole-plot / 
between-subjects factors) and repeated measures (sub-plot / within-subjects 
factors) can be included by splitting the indices $i$ and $\ell$, respectively. 
Also in this setting we can define adequate model parameters on the marginals 
$X_{i\ell 1}\sim F_{i\ell}$. In particular, those are given by the relative 
effect {$p_{i\ell}$} of the distribution of group $i$ at time 
$\ell$ with respect to the unweighted pooled distribution function $G = 
\frac{1}{dt} \sum_{i=1}^d \sum_{\ell=1}^t F_{i\ell}$ { by} \sep 
\bqan
\label{releff rm}
p_{i\ell} &=& \int G dF_{i\ell} \quad \ig d; \  \ell=1,\dots,t.
\eqan
This relative effect $p_{i\ell}$ can also be written as the mean 
$p_{i\ell} = \overline{w}_{\cdot\cdot i\ell}$ of the relative marginal effects 
$w_{rsi\ell} = \int F_{rs} d F_{i\ell}, 1\leq i,r\leq d, 1\leq s,\ell\leq t$. 
Collecting all $p_{i\ell}$ in a vector $\vp =(p_{11},p_{12},\dots,p_{dt})'$ 
{ the} linear hypotheses of interest can be written as $H_0^p: 
\vC\vp = \vnull$ using an adequate hypothesis matrix $\vC$ in the 
same way as in the linear models setting. Inference methods for testing more 
restrictive null hypotheses formulated in terms of distribution functions have 
been developed by Akritas and Brunner (1997) and  Brunner et al. (1999).

{\color{black} A factorial structure on the groups or repeated measures is easily obtained in this setup by splitting 
the indices $i$ or $\ell$ into sub-indices $i', i'', \ldots$ or $\ell', \ell'', \ldots$, 
respectively. Thus, higher-way layouts with repeated measures or longitudinal 
data are covered by the general model defined in (\ref{model rm}).}

For testing $H_0^p$, estimates for the effects $p_{i\ell}$ are obtained as in 
Section~\ref{mod} by substituting the distribution functions $F_{i\ell}(x)$ in
\eqref{releff rm} with their empirical counterparts $\hF_{i\ell}(x) = 
\frac1{n_i} \sumk {n_i} c(x-X_{i\ell k})$ resulting in 
\bqan \label{releffest rm}
 \hp_{i\ell} &=& \int \hG d \hF_{i\ell}.
\eqan

Thus, an estimator for the vector $\vp$ is given by $\vwhp =(\hp_{11}, 
\hp_{12},\dots,\hp_{dt})'$ and its asymptotic behaviour can be studied similar 
to Section~3. In particular, an application of the asymptotic equivalence theorem
and the Cramer-Wold-Device shows that $\sqrt{N} (\vwhp - \vp)$ possesses an asymptotic 
multivariate normal distribution with mean $\vnull$ and an unknown covariance 
matrix $\vV$. Its rather complex form is given in the supplementary material 
where we also introduce a consistent rank-based estimator of $\vV$. {
This allows us to develop ANOVA-type tests for $H_0^p:\vC\vp = \vnull$} 
following the same steps as in Section~4. For ease of presentation we will 
apply these methods with more details in a future paper, where we also 
investigate their finite sample behaviour.
 
\section{Specific Designs} \label{spd}

In this section we apply the results derived in the previous sections to some frequently used specific designs, where the hypotheses are formulated in terms of the nonparametric effects $p_i = \int G\ dF_i$ as defined in \eqref{releff}. 
These unweighted effects, proposed by Brunner and Puri (2001, Section~3.2) and later considered in detail by Domhof (2001) as well as Gao and Alvo (2005) and Gao et al. (2008), 
can be regarded as nonparametric treatment effects which are defined by the distributions in the designs. 
Thus, they are fixed model-based constants in contrast to the effects $r_i$ defined in the introduction and first discussed by Kruskal and Wallis (1952). 
Since these $r_i$ depend on the sample sizes by definition, it is not reasonable to formulate hypotheses about 
these sample size dependent quantities $r_i$. 
The nonparametric effects $p_i$ do not have these drawbacks. Moreover, they can be interpreted as effects of 
$F_i$ with respect to $G= \frac{1}{d}\sum_{\ell=1}^d F_\ell$, i.e.
\bqa
p_i &=& P(Z_G < X_{i1}) + \tfrac12 P(Z_G = X_{i1}),
\eqa
where $Z_G\sim G$ is independent of $X_{i1} \sim F_i$. 
{\color{black} We note that these effects $p_i$ correspond to the means in the classical 
homoscedastic ANOVA models where the means represent the centers of gravity of the distributions. 
More general, the $p_i$ describe a tendency to larger or smaller values of observations from group $i$ with respect to the mean distribution $G$.
}

This motivates us to formulate hypotheses $H_0^p(\vT): \vT \vp =\vnull$ {\color{black}(as in defined Section~\ref{stat})} based on 
these nonparametric effects in general factorial designs. It is technically 
managed as in classical linear model theorems by splitting up the indices $\ig 
d$ into sub-indices $i_1,i_2,\dots$ according to the factorial structure of the
design. In the one-way layout, for example, where the factor $A$ has $\ig a$ 
levels we have $X_{ik} \sim F_i, \ig a$. Here the null hypothesis of equal 
treatment effects is formulated as
\bqa
H_0^p(A): \vP_a \vp &=& \vnull,
\eqa
where $\vP_a = \vI_a - \tfrac{1}{a} \vJ_a$ denotes the $a$-dimensional centring
matrix. In this case, the statistic ${\color{black}Q_N}(\vT)$ in \eqref{ATS} reduces to
\bqa
{\color{black}Q_N}(\vP_a) &=& \frac{N }{\tr(\vP_a \vwhV)} \vwhp' \vP_a \vwhp \ = \ 
\frac{N}{\tr(\vP_a \vwhV)} \sumi {a} (\whp_i - \tfrac{1}{2})^2 
\eqa
by noting that $\olp_{\cdot} = \tfrac{1}{a} \sum_{i=1}^a \int G \ dF_i = \int G 
dG = \frac{1}{2}$. Under $H_0^p(A): \vP_a\ \vp =\vnull$, the statistic 
${\color{black}Q_N}(\vP_a)$ can be approximated by one of the three methods described in 
Section~\ref{stat}, where $\vT$ is  replaced by $\vP_a$. 

In a two-way layout with two crossed factors $A$ and $B$, with levels $\ig a$ 
and $\jg b$, respectively, it is only assumed that the observations 
\bqa
 X_{ijk} & \sim & F_{ij},\quad \ig a; \ \jg b; \ \kg n_{ij}
\eqa
are independent and non-constant. Then using the mean distribution function $G 
=\frac{1}{ab}\sumi a \sumj b F_{ij}$ the \np\ effects are written as $p_{ij} = 
\int G \ dF_{ij}$ and are collected in the vector $\vp=(p_{11},\ldots 
p_{ab})'$. {\color{black} Their interpretation is as follows: 
If $p_{ij}\leq p_{rs}$ the observations under factor combination $(i,j)$ tend to result in smaller values as the observations under factor combination $(r,s)$. 
Moreover, another interpretation can be given in terms of additive effects by 
using a decomposition of the distribution functions as in Akritas and Arnold (1994) and 
the supporting information in de Neve and Thas (2015){\color{black}: Writing 
$G= \olF_{\cdot \cdot} = \frac{1}{ab}\sum_{i=1}^a\sum_{j=1}^b F_{ij}$, $A_i=\olF_{i\cdot}-G = \frac{1}{b}\sum_{j=1}^b F_{ij} - G$, 
$B_j=\olF_{\cdot j}-G = \frac{1}{a}\sum_{i=1}^a F_{ij} - G$ and $(AB)_{ij}=F_{ij}-\olF_{i\cdot}-\olF_{\cdot j}+G$ we have $F_{ij}=G+A_i+B_j+(AB)_{ij}$. 
Plugging this into the definition of the nonparametric effect results in an additive effects representation as in classical linear models
\bqan\label{eq:decomposition}
 p_{ij} = \int G d F_{ij}  =   
  \frac{1}{2} + {\int G d A_i} + {\int G d B_j} + {\int G d(AB)_{ij}}
   \equiv 
  \frac{1}{2} + {\alpha_i} + {\beta_j} + {(\alpha\beta)_{ij}}
  \eqan
with $\sum_{i=1}^a \alpha_i= \sum_{j=1}^b \beta_j =0$, $\sum_{i=1}^a (\alpha\beta)_{ij}=0$ for all $j=1, \ldots, b$ and $\sum_{j=1}^b (\alpha\beta)_{ij}=0$ for all $i=1, \ldots, a$, see the supplement for details. 
Here we can, e.g., rewrite the additive effect $\beta_j$ as
$$
 \beta_j = \int G dF_{\cdot j} -\frac{1}{2} = P(Z_G< Z_{bj}) + \frac{1}{2}P(Z_G = Z_{bj}) - \frac{1}{2}
$$
to gain the following interpretation: 
If $\beta_j>0$ it is more likely that a randomly selected observation 
$Z_{bj}\sim \olF_{\cdot j}$ from the mean distribution $\olF_{\cdot j}$ under level $j$ of factor $B$ is larger than 
a randomly selected observation $Z_G$ from the mean distribution $G$.
}

Finally, nonparametric hypotheses {\color{black} for  \np\ effects} are formulated as {\color{black} in classical linear models via contrast matrices as}
\bqa
H_0^p(A): \vT_A\ \vp &=& \vP_a \otimes \tfrac{1}{b}{\color{black}\vJ_b}\ \vp \ =\ \vnull 
\qquad \text{(no nonparametric main effect A)} \\
H_0^p(B): \vT_B \ \vp &=& \tfrac{1}{a}{\color{black}\vJ_a} \otimes \vP_b\ \vp\ =\ \vnull 
\qquad \text{(no nonparametric main effect B)} \\
H_0^p(AB): \vT_{AB} \ \vp &=& \vP_a \otimes \vP_b\ \vp \ =\ \vnull  \qquad 
\text{(no nonparametric interaction effect AB)}.
\eqa
{\color{black}  With the decomposition \eqref{eq:decomposition} the above hypotheses can be written in a more popular way, e.g. 
$H_0^p(A): \vT_A\ \vp = \vnull \iff \alpha_1=\dots =\alpha_a=0$. }

{\color{black} Moreover, similar contrast matrices can be used in the context of 
repeated measures designs, see {\color{black} e.g., Section~1.3 in Akritas and 
Brunner (1997)} for related hypotheses formulated in terms of distribution 
functions.

We note that similar hypotheses have been discussed by Boos and Brownie (1992) 
in the special case of an $(a\times2)$-design. The statistics can again be 
derived by plugging in $\vT_R$ for $\vT$ in \eqref{ATS}, $R\in\{A,B,AB\}$. }
{\color{black} Also Akritas et al. (1997) have discussed the meaning and 
interpretation of general nonparametric effects in terms of the distribution 
functions in detail. The nonparametric effect $p_{ij}$ is simply a measure of 
an overlap of the distribution function $F_{ij}(x)$ with the mean distribution 
function $G(x)$ and thus is to be understood in the same line as a 
nonparametric effect based on the distribution functions.} \\

\section{Simulations} \label{sim}
Next we investigate the small sample properties of the three statistical tests $\widehat{\varphi}_N$, $\widetilde{\varphi}_N$, and $\varphi_N$ based on the ANOVA-type statistic ${\color{black}Q_N}(\vT)$ in \eqref{ATS} with the three approximations \eqref{ATSlambda}, \eqref{eq: df2}, and \eqref{whf1}, respectively, within extensive simulation studies with regard to their
\bit
\item[(a)] maintenance of the preassigned type I error level ($\alpha = 5\%$) under the hypothesis $H_0^p(\vT): \vT \vp = \vnull$ and 
\item[(b)] their powers to detect specific alternatives.
\eit

All simulations were performed using {\it R} (version 2.15.0, R Development Core Team, 2010) with $nsim=10,000$ simulation runs for each setting. 
The distribution of $\whQ(\vT)$ given in (\ref{ATSlambda}) was approximated using $n_{MC}=10,000$ Monte-Carlo runs, and the critical values 
were estimated from this distribution. Hereby, the eigenvalues of the matrix $\vT\vwhV_N$ were computed with the base R-function
\textit{eigen}. 

In order to compare the newly developed methods with {\color{black} other procedures we first} 
restrict our considerations to the one-way layout (balanced and unbalanced) with $a=4$ 
independent treatment groups, and by using both symmetric and skewed distributions. 
{\color{black} In this set-up the above procedures test the null hypothesis $H_0^p: p_1=p_2=p_3=p_4$. 
As competi{\color{black}t}ors the classical Kruskal-Wallis rank test and two Wald-type tests are considered: 
The test ${\color{black}\varrho_N} = \veins\{{\color{black}W_N}(\vC) > \chi^2_{1-\alpha;r({\color{black}\vwhM_N})}\}$ based on the WTS given in {\color{black}\eqref{WTS}} and a related 
test in a Wald-type statistic for a probabilistic index model (PIM, Thas et al., 2012) using a sandwich-type covariance matrix estimator, say $\vwhS$, 
and weighted rank estimators for the PIM effects, say $\vwhalpha$, instead of $\vwhV_N$ and $\vwhp$, respectively, and a $\chi^2$-quantile with estimated degrees of freedom given by $r(\vC\vwhS\vC')$. 
The latter is motivated from the considerations in de Neve and Thas (2015) {\color{black}and denoted as DTS}. 
We note that it is a test for the related null hypothesis $H_0^\alpha: \alpha_1=\dots=\alpha_4$ formulated in terms of the weighted 
PIM effects $\alpha_i$ (see Equation (4) in de Neve and Thas, 2015, for its explicit definition) which is equal to $H_0^p$ in the balanced case. 
The ingredients of the test statistic were calculated as described in the supplementary material of de Neve and Thas (2015) with the {\it R} package {PIM} (Version 1.1.5.6). 
Moreover, note that the Kruskal-Wallis test has been developed for testing the more restrictive null hypothesis $H_0^F: F_1=F_2=\dots=F_a$ formulated in terms of the distribution functions.
}

Symmetrically distributed data was generated from the model 
\bqa X_{ik} = \mu_i  +\sigma_{i} \epsilon_{ik}, \quad i=1,\ldots,a; \sep k=1,\ldots,n_{i},
\eqa
where the random error terms
\bqa
\epsilon_{ik}= \frac{\wtep_{ik} - E(\wtep_{i1})}{\sqrt{\Var(\wtep_{i1})}}
\eqa 
were generated from different standardized symmetric distributions, i.e., the random variables $\wtep_{ik}$ were generated from 
standard normal {\color{black} or the}  double exponential distribution, respectively. 
 Skewed data was generated from log-normal-distributions by $X_{ik} = \exp(\eta_{ik})$, where $\eta_{ik}\sim N(0,\sigma_i^2)$ and possibly different variances $\sigma_i^2$. Note that the null hypothesis $H_0^p: \vP_a\vp=\vnull$ holds in both cases, because of the symmetry and the monotonicity of the exponential function.

A major assessment criterion for the accuracy of the methods is their behavior when different sample sizes and variances are combined, i.e. when increasing sample sizes are combined with increasing variances (\textit{positive pairing}) or with decreasing variances (\textit{negative pairing}) 
(see Pauly et al., 2015a). We consider balanced situations with sample size vector $\vn_1=(n_1,n_2,n_3,n_4)=(5, 5, 5, 5)$ and unbalanced situations with sample size vector $\vn_2=(n_1,n_2,n_3,n_4)=(10,20,30,40)$, respectively. The scaling vector $\vsigma=(\sigma_1,\sigma_2,\sigma_3,\sigma_4)$ was chosen from $(1,1,1,1), (1,\sqrt{2},2,\sqrt{5})$ or $(\sqrt{5},2,\sqrt{2},1)$, respectively. In order to investigate the behavior of the tests when the sample sizes increase, a constant $m \in \{5,10,20,25\}$ was added to each component of the vectors $\vn_1$ and $\vn_2$, i.e. $\vn_i + m\veins_4' = (n_1 + m, n_2 + m, n_3 + m, n_4 + m), i=1,2$. The different simulation settings are summarized in Table~\ref{SimSetting1}. 

\begin{table}[h]
\caption{Simulated one-way layout with $a=4$ samples, where $m \in \{0, 5, 10, 20, 25\}$ and $\vn_1=(5, 5, 5, 5)$, and $\vn_2=(10,20,30,40)$. } \label{SimSetting1}
\bcen
\begin{tabular}{clll} \hline
 Setting & Sample Size & Scaling Factors & {\color{black} Meaning} \\\hline
 1 & $\vn=\vn_1+m\veins_4'$ & $\vsigma=(1,1,1,1)$ & Balanced  homoscedastic \\
 2 & $\vn = \vn_2+m\veins_4'$ & $\vsigma=(1,1,1,1)$ & Unbalanced  homoscedastic \\
 3  &$\vn = \vn_1+m\veins_4'$ & $\vsigma=(1,\sqrt{2},2,\sqrt{5})$ & Balanced  heteroscedastic \\
 4  &$\vn = \vn_2+m\veins_4'$ & $\vsigma=(1,\sqrt{2},2,\sqrt{5})$ &  {\color{black} Unbalanced  heteroscedastic (}Positive Pairing{\color{black})}\\
 5  &$\vn = \vn_2+m\veins_4'$ & $\vsigma=(\sqrt{5},2,\sqrt{2},1)$ & {\color{black} Unbalanced  heteroscedastic (}Negative Pairing{\color{black})}\\\hline \hline
\end{tabular}
\ecen
\end{table}

{\color{black}
Because of space limitation, we only display the results of the two balanced settings 1 and 3 in Tables~\ref{Table 1} and \ref{Table 3} below. 
The simulation results of the other settings 2, 4, and 5 are listed in Section 4 (More Simulation Results) of the supplementary material. 
For the homoscedastic balanced case (Table~\ref{Table 1})
}
the Kruskal-Wallis test controls the nominal type-1 
error level ($\alpha=5\%$) very satisfactorily for all investigated  
distributions. This result is not surprising, because {\color{black} in this case} the hypothesis $H_0^F$ holds. 
{\color{black}  The {\color{black}DTS and the WTS} tend to be highly liberal for small sample sizes ($n_i\leq 15$). With increasing sample sizes the liberality of 
both tests slowly decreases. However, even for the scenarios with larger sample sizes their type-$I$-error control is not acceptable. 
A similar behaviour of such Wald-type statistics has been observed for various models, see e.g. Vallejo et al. (2010), Pauly et al. (2015a) or DiCiccio and Romano (2015). 
The behaviour of the ANOVA-type tests is different.} 
For smaller sample sizes ($n_i \leq 25$) both 
the tests $\widehat{\varphi}_N$ and $\widetilde{\varphi}_N$ tend to result in 
more or less liberal conclusions. In case of 'extreme' small samples ($n_i =5$), 
the estimated type-1 error level is about $8\%$. The ANOVA-type test $\varphi_N$
based on the $F$-approximation of the statistic ${\color{black}Q_N}(\vT)$ controls the type-1 
error level even for extreme small sample sizes and under all investigated 
distributions.

Next we comment on the {\color{black}balanced} heteroscedastic setting {\color{black}3 displayed} in Table~\ref{Table 3}. 
Note, that here $H_0^F$ is violated and only $H_0^p$ holds true. 
The Kruskal-Wallis test tends to over reject the null hypothesis under normality. Under the assumption of log-normal or double exponential distributions, 
this statistic fairly controls the type-1 error level. 
{\color{black}Again, both of the Wald-type tests {\color{black}(DTS and WTS)} are liberal,} the ANOVA-type tests $\widehat{\varphi}_N$ and 
$\widetilde{\varphi}_N$ tend to be fairly liberal, while the test $\varphi_N$ 
controls the type-1 error level at best.

{\color{black} Summarizing the above simulation results it turns out that the ANOVA-type test $\varphi_N$ based on the $F$-approximation of the statistic 
${\color{black}Q_N}(\vT)$ turns out to control the type-$I$ error rate at best in all considered case{\color{black}s here. The same remark also holds 
for the other simulation settings shown in the supplementary material. Thus, the ANOVA-type test $\varphi_N$} is recommended for practical applications. 
}

Next we investigate the powers of the procedures to detect certain alternatives.
Data is generated by $X_{ik}\sim N(\mu_i,1), i=1,\ldots,4; k=1,\ldots,n$, and 
sample sizes $n_i \equiv n \in \{15,20\}$.  {\color{black} Due to their liberal behaviour in all investigated settings we do not consider both Wald-type tests in the power simulations and subsequent considerations regarding factorial designs.}
We consider two types of 
alternatives:
\ben
\item[(1)] \sep $\vmu=(\mu_1,\mu_2,\mu_3,\mu_4)' = (0,0,0,\delta)'$ {\color{black}\sep -- \sep (one-point alternative),} \\
\item[(2)] \sep $\vmu=(\mu_1,\mu_2,\mu_3,\mu_4)' = (\frac{\delta}4, 
\frac{\delta}2, \frac{3 \delta}4, \delta)'$ {\color{black}\sep -- \sep (increasing-trend alternative).}
\een

In both cases, $\delta$ is increased as $\delta=0,0.1,\ldots,1.6$. The 
simulation results are displayed in Table~\ref{Power}. It can be seen that both
the powers of the Kruskal-Wallis test (KW) and the ANOVA-type test $\varphi_N$ 
using the $F$-approximation are very likely and none of the two procedures is 
superior to the other. Furthermore, both the powers of the ANOVA-type tests 
$\widehat{\varphi}_N$ and $\widetilde{\varphi}_N$ are slightly higher than 
those of the Kruskal-Wallis test and $\varphi_N$ which may be explained by 
their slight{\color{black}ly} liberal behaviour. 
We have also run simulations for non-normal data where we obtained similar
results (not presented here).\\


\begin{table}[H]{\color{black}
\caption{Type-I error ($\alpha=5\%$) simulations of the Kruskal-Wallis test (KW), the two Wald-type tests in the test statistics WTS and the test statistic of De Neve and Thas (DTS) and the 
three different ANOVA-type tests $\widehat{\varphi}_N$, $\widetilde{\varphi}_N$, and $\varphi_N$ using the distributional approximations as given in \eqref{ATSlambda}, \eqref{eq: df2}, and \eqref{whf1} 
under Setting~1 as described in Table~\ref{SimSetting1}. }\label{Table 1}
{\footnotesize\bcen
\begin{tabular}{cccccccc}\hline
Distribution   & Sample Sizes                                     & KW              &  DTS              & WTS              &  $\widehat{\varphi}_N$           &  $\widetilde{\varphi}_N$ & ${\varphi}_N$\\\hline
DExp	&	5	5	5	5	        &	0.0367	&	0.3460	&	0.2216	&	0.0743	&	0.0751	&	0.0348	\\
DExp	&	10	10	10	10	&	0.0428	&	0.1868	&	0.1265	&	0.0621	&	0.0628	&	0.0460	\\
DExp	&	15	15	15	15	&	0.0463	&	0.1324	&	0.0922	&	0.0610	&	0.0611	&	0.0477	\\
DExp	&	25	25	25	25	&	0.0456	&	0.0862	&	0.0752	&	0.0541	&	0.0552	&	0.0471	\\
DExp	&	30	30	30	30	&	0.0499	&	0.0888	&	0.0706	&	0.0570	&	0.0570	&	0.0510	\\\hline
LogNor	&	5	5	5	5	        &	0.0376	&	0.3370	&	0.2233	&	0.0789	&	0.0803	&	0.0377	\\
LogNor	&	10	10	10	10	&	0.0461	&	0.1762	&	0.1183	&	0.0663	&	0.0665	&	0.0476	\\
LogNor	&	15	15	15	15	&	0.0475	&	0.1256	&	0.0938	&	0.0614	&	0.0619	&	0.0493	\\
LogNor	&	25	25	25	25	&	0.0433	&	0.0904	&	0.0756	&	0.0506	&	0.0512	&	0.0442	\\
LogNor	&	30	30	30	30	&	0.0498	&	0.0854	&	0.0687	&	0.0552	&	0.0555	&	0.0509	\\\hline
Normal	&	5	5	5	5	        &	0.0348	&	0.3354	&	0.2223	&	0.0772	&	0.0784	&	0.0361	\\
Normal	&	10	10	10	10	&	0.0442	&	0.1766	&	0.1229	&	0.0631	&	0.0646	&	0.0469	\\
Normal	&	15	15	15	15	&	0.0472	&	0.1268	&	0.0941	&	0.0614	&	0.0616	&	0.0491	\\
Normal	&	25	25	25	25	&	0.0466	&	0.0910	&	0.0744	&	0.0544	&	0.0547	&	0.0480	\\
Normal	&	30	30	30	30	&	0.0498	&	0.0806	&	0.0691	&	0.0548	&	0.0549	&	0.0509	\\\hline
\end{tabular}
\ecen}}
		\end{table}

\begin{table}[H]{\color{black}
\caption{Type-I error ($\alpha=5\%$) simulations of the Kruskal-Wallis test (KW), the two Wald-type tests in the test statistics WTS and the test statistic of De Neve and Thas (DTS) and the 
three different ANOVA-type tests $\widehat{\varphi}_N$, $\widetilde{\varphi}_N$, and $\varphi_N$ using the distributional approximations as given in \eqref{ATSlambda}, \eqref{eq: df2}, and \eqref{whf1} 
under Setting~3 as described in Table~\ref{SimSetting1}.}\label{Table 3}
{\footnotesize\bcen
\begin{tabular}{cccccccc}\hline
Distribution   & Sample Sizes                                     & KW              &  DTS              & WTS              &  $\widehat{\varphi}_N$           &  $\widetilde{\varphi}_N$ & ${\varphi}_N$\\\hline
DExp	&	5	5	5	5	        &	0.0547	&	0.3316	&	0.2232	&	0.0838	&	0.0842	&	0.0419	\\
DExp	&	10	10	10	10	&	0.0592	&	0.1658	&	0.1219	&	0.0705	&	0.0719	&	0.0507	\\
DExp	&	15	15	15	15	&	0.0626	&	0.1256	&	0.0910	&	0.0641	&	0.0646	&	0.0515	\\
DExp	&	25	25	25	25	&	0.0629	&	0.0908	&	0.0768	&	0.0576	&	0.0580	&	0.0497	\\
DExp	&	30	30	30	30	&	0.0625	&	0.0866	&	0.0665	&	0.0564	&	0.0573	&	0.0506	\\\hline
LogNor	&	5	5	5	5	        &	0.0399	&	0.3416	&	0.2200	&	0.0792	&	0.0801	&	0.0372	\\
LogNor	&	10	10	10	10	&	0.0445	&	0.1630	&	0.1166	&	0.0628	&	0.0635	&	0.0460	\\
LogNor	&	15	15	15	15	&	0.0471	&	0.1310	&	0.0961	&	0.0600	&	0.0604	&	0.0483	\\
LogNor	&	25	25	25	25	&	0.0514	&	0.0874	&	0.0753	&	0.0582	&	0.0582	&	0.0510	\\
LogNor	&	30	30	30	30	&	0.0492	&	0.0804	&	0.0691	&	0.0533	&	0.0538	&	0.0482	\\\hline
Normal	&	5	5	5	5	&	0.0572	&	0.3294	&	0.2281	&	0.0847	&	0.0858	&	0.0398	\\
Normal	&	10	10	10	10	&	0.0669	&	0.1646	&	0.1277	&	0.0733	&	0.0739	&	0.0520	\\
Normal	&	15	15	15	15	&	0.0654	&	0.1264	&	0.0983	&	0.0630	&	0.0645	&	0.0521	\\
Normal	&	25	25	25	25	&	0.0673	&	0.0862	&	0.0738	&	0.0577	&	0.0585	&	0.0515	\\
Normal	&	30	30	30	30	&	0.0662	&	0.0818	&	0.0695	&	0.0562	&	0.0566	&	0.0494	\\\hline
\end{tabular}
		
		\ecen}}
\end{table}

\begin{sidewaystable}[H]
\caption{Comparison of the power (type-1 error level $\alpha=5\%$) for the Kruskal-Wallis test (KW) and the three ANOVA-type tests $\widehat{\varphi}_N$, $\widetilde{\varphi}_N$, and ${\varphi}_N$ based on the approximations described in \eqref{ATSlambda}, \eqref{eq: df2}, and \eqref{whf1} for two different shift alternatives and two different balanced designs using normal data with homogeneous variances. }\label{Power}
{\footnotesize\bcen
\begin{tabular}{cccccccccccccccccccccc}\hline\\
& \multicolumn{9}{c}{Alternative $\vmu=(0,0,0,\delta)$} && \multicolumn{9}{c}{Alternative $\vmu=(\delta,\frac{\delta}{2},3\frac{\delta}{4},\frac{\delta}{4})'$}\\\hline\\
&\multicolumn{4}{c}{$n_i \equiv n=15$} && \multicolumn{4}{c}{$n_i \equiv n=20$} &&  \multicolumn{4}{c}{$n_i \equiv n=15$} && \multicolumn{4}{c}{$n_i \equiv n=20$}\\\hline
 $\delta$  & KW  &  $\widehat{\varphi}_N$    &  $\widetilde{\varphi}_N$ & ${\varphi}_N$ && KW  &  $\widehat{\varphi}_N$    &  $\widetilde{\varphi}_N$ & ${\varphi}_N$ && KW  &  $\widehat{\varphi}_N$    &  $\widetilde{\varphi}_N$ & ${\varphi}_N$ && KW  &  $\widehat{\varphi}_N$    &  $\widetilde{\varphi}_N$ & ${\varphi}_N$\\\hline
 0.0  & 0.0467 & 0.0603  & 0.0615 & 0.0491 && 0.0472 & 0.0587 & 0.0587 & 0.0483 && 0.0406 &0.0549 & 0.0552 &0.0434  &&0.0487 &0.0580 & 0.0586 &0.0509  \\
 0.1  & 0.0517 & 0.0638  & 0.0639 & 0.0535 && 0.0515 & 0.0601 & 0.0605 & 0.0518 && 0.0494 &0.0630 & 0.0637 &0.0512  &&0.0498 &0.0599 &0.0599 &0.0505  \\
 0.2  & 0.0738 & 0.0913  & 0.0919 & 0.0764 && 0.0785 & 0.0904 & 0.0901 & 0.0803 && 0.0574 &0.0725 & 0.0735 &0.0606  &&0.0641 &0.0755 &0.0753 &0.0657\\
 0.3  & 0.1008 & 0.1196  & 0.1205 & 0.1046 && 0.1293 & 0.1463 & 0.1469 & 0.1323 && 0.0698 &0.0855 & 0.0867 &0.0721  &&0.0761 &0.0892 &0.0897 &0.0784 \\
 0.4  & 0.1548 & 0.1842  & 0.1854 & 0.1583 && 0.1996 & 0.2217 & 0.2215 & 0.2047 && 0.0902 &0.1121 & 0.1122 &0.0930  &&0.1037 &0.1199 &0.1199 &0.1056\\
 0.5  & 0.2208 & 0.2523  & 0.2535 & 0.2249 && 0.3072 & 0.3368 & 0.3383 & 0.3123 && 0.1125 &0.1364 & 0.1358 &0.1159  &&0.1376 &0.1564 &0.1571 &0.1415\\
 0.6  & 0.3179 & 0.3601  & 0.3610 & 0.3227 && 0.4184 & 0.4554 & 0.4546 & 0.4241 && 0.1475 &0.1731 & 0.1746 &0.1518  &&0.1874 &0.2092 &0.2107 &0.1912\\
 0.7  & 0.4136 & 0.4588  & 0.4599 & 0.4230 && 0.5525 & 0.5877 & 0.5880 & 0.5602 && 0.1854 &0.2164 & 0.2178 &0.1901  &&0.2478 &0.2754 &0.2776 &0.2532\\
 0.8  & 0.5316 & 0.5767  & 0.5786 & 0.5402 && 0.6830 & 0.7118 & 0.7125 & 0.6893 && 0.2335 &0.2723 & 0.2729 &0.2409  &&0.3173 &0.3474 &0.3491 &0.3230\\
 0.9  & 0.6430 & 0.6849  & 0.6870 & 0.6500 && 0.7936 & 0.8186 & 0.8198 & 0.7989 && 0.3000 &0.3417 & 0.3434 &0.3080  &&0.4014 &0.4345 &0.4366 &0.4091\\
 1.0  & 0.7421 & 0.7786  & 0.7785 & 0.7486 && 0.8777 & 0.8962 & 0.8965 & 0.8815 && 0.3686 &0.4125 & 0.4153 &0.3763  &&0.4900 &0.5244 &0.5277 &0.4982\\
 1.1  & 0.8292 & 0.8579  & 0.8589 & 0.8358 && 0.9364 & 0.9476 & 0.9479 & 0.9389 && 0.4307 &0.4777 & 0.4780 &0.4385  &&0.5617 &0.5917 &0.5914 &0.5688\\
 1.2  & 0.8968 & 0.9185  & 0.9180 & 0.9015 && 0.9687 & 0.9758 & 0.9759 & 0.9712 && 0.5074 &0.5494 & 0.5503 &0.5139  &&0.6485 &0.6807 &0.6815 &0.6546\\
 1.3  & 0.9410 & 0.9549  & 0.9549 & 0.9433 && 0.9873 & 0.9907 & 0.9906 & 0.9877 && 0.5778 &0.6222 & 0.6228 &0.5862  &&0.7338 &0.7624 &0.7654 &0.7398\\
 1.4  & 0.9703 & 0.9767  & 0.9769 & 0.9717 && 0.9950 & 0.9967 & 0.9966 & 0.9953 && 0.6566 &0.6982 & 0.6991 &0.6662  &&0.8003 &0.8238 &0.8240 &0.8059\\
 1.5  & 0.9847 & 0.9887  & 0.9886 & 0.9856 && 0.9983 & 0.9986 & 0.9986 & 0.9984 && 0.7175 &0.7538 & 0.7558 &0.7260  &&0.8580 &0.8771 &0.8777 &0.8621\\
 1.6  & 0.9922 & 0.9945  & 0.9943 & 0.9926 && 0.9994 & 0.9996 & 0.9996 & 0.9994 && 0.7820 &0.8159 & 0.8157 &0.7880  &&0.9011 &0.9138 &0.9144 &0.9036\\\hline
\end{tabular}
\ecen}
\end{sidewaystable}

{\color{black}
\section{Software and Analysis of the Data Example} \label{exa}

In order to provide freely available software for data analysis and educational purposes we implemented 
an {\it R} software package called {\bf rankFD} for rank based analysis of independent observations in factorial designs. 
For a user-friendly implementation it is equipped with a graphical user interface. 
The package contains the ANOVA-type-p-test (who turned out to be the best in our simulation study) for making inference in one-, two- or arbitrary higher-way layouts 
as well specific nested designs. Furthermore, all test procedures for testing the hypothesis $H_0^F$ formulated in terms of the distribution functions are implemented. 
Besides of a descriptive overview it also provides $p$-values and confidence intervals for the main treatment effects along with plotting options. 
The \textit{R} package will be updated frequently.
The {\it R}-package is freely available at CRAN. Here it has been exemplified for analysing the motivating data example  described in the Introduction.

{\color{black}
The statistics and $p$-values for testing the main effects $A$ ({\it food
condition}) and $B$ ({\it treatment}) as well as the interaction $AB$ between
the food condition and the treatment are listed in Table~\ref{Table Ex}. 
}

\begin{table}[H]
\caption{Analysis of the data example with the ANOVA-type$-p$-test $\varphi_N$
{\color{black}given in Theorem~\ref{astests}(b)(3).} The value of the test statistic $Q_N(\vT)$ is compared with the quantile of an
$F$-distribution with estimated degrees of freedom $\whf_1$ and $\whf_2$.}
\label{Table Ex}
{\small
\bcen
\begin{tabular}{lrccr}
Factor & Statistic & $\whf_1$ & $\whf_2$ & $p$-value \\\hline
Food {\color{black}Condition}& 42.450 & 1 & 26.492 & $< 0.0001$ \\
Treatment & 33.191 & 1 & 26.492 & $< 0.0001$ \\
Interaction &  1.868 & 1 & 26.492 & $0.1832$ \\
\hline
\end{tabular}
\ecen
}

\end{table}

It {\color{black}appears} from Table~\ref{Table Ex} that both the factors \textit{Food} as
well as \textit{Treatment} have a significant impact on the numbers of
leucocytes at 5\% level. The data do not provide {\color{black}any} evidence {\color{black}for an interaction between the treatment and the food condition.

Point estimates of the nonparametric treatment effects $p_{ij} = \int G dF_{ij}$ for each drug $\times$ food
combination are computed. The index $i$ refers to the factor $A$ (food
condition: $i=1$, normal food; $i=2$, reduced food) while the second index $j$ refers to the factor $B$ (treatment: $j=1$, placebo; $j=2$, drug). Also
two-sided (range preserving) 95\%-confidence intervals for the $p_{ij}$ are computed as given in \eqref{eq:CIg} where the logit
transformation $g(x) = \log(x / (1-x))$ has been used. The results are listed in
Table~\ref{TableDescriptive}.

\begin{table}[H]\color{black}
\caption{Estimates and 95\%-confidence intervals for the nonparametric treatment effects $p_{ij} = \int G dF_{ij}$ in the leucocytes trial. The index
$i$ refers to the food condition while the index $j$ refers to the treatment.
The range-preserving limit are obtained by the logit-transformation $g(x) =
\log(x / (1-x))$.}
\label{TableDescriptive}
\bcen
\begin{tabular}{llclll}\hline
\mc{2}{c}{Factor Level Combination} & Sample Size & \mc{1}{c}{Effect} &
\mc{2}{c}{95\%-Confidence Limits}\\ \hline
\mc{1}{c}{Food Condition} & \mc{1}{c}{Treatment} & $n_{ij}$ &
\mc{1}{c}{$\hp_{ij}$} & Lower & Upper \\ \hline
$i=1$ \ - \ Normal & $j=1$ \ - \ Placebo & 10 & 0.460 & 0.355 & 0.568 \\ 
$i=1$ \ - \ Normal & $j=2$ \ - \ Drug & 10 & 0.855 & 0.818 & 0.885   \\
$i=2$ \ - \ Reduced & $j=1$ \ - \ Placebo & 10 & 0.209 & 0.140 & 0.301 \\
$i=2$ \ - \ Reduced & $j=2$ \ - \ Drug & 10 &  0.476 & 0.375 & 0.579 \\ \hline
\end{tabular}
\ecen
\end{table}

The estimated effect $\hp_{21} = 0.209$ for the reduced food under placebo means that the
observations from $F_{21}$ tend to be smaller than those from the mean
distribution $G=\frac{1}{4}\sum_{i,j=1}^2 F_{ij}$, or more precisely, the probability that a randomly selected 
observation $Z$ from the mean distribution $G$ is smaller than a randomly selected
observation $X_{21}$ from $F_{21}$ equals $0.209$. Similarly, the estimated effect $\hp_{12} =
0.855$ for the normal food under the drug means that the observations from
$F_{12}$ tend to be larger than those from the mean distribution $G$. 
We note that the confidence intervals for Placebo and Drug do not overlap within each food condition which may be interpreted that the drug is effective in both cases.
}

\section{Discussion} \label{dis}

Rank methods for the analysis of factorial designs denote a substantial and important area in statistical research and applications. Both the 
Wilcoxon-Mann-Whitney test and the Kruskal-Wallis test can be viewed as 
one of the most frequently applied nonparametric methods. Furthermore, these procedures have been generalized for the analysis of factorial layouts by several authors. However, up to now, all of these 
{\color{black} are only worked out} to test hypotheses being formulated in terms of the distribution functions, i.e. 
$H_0^F(\vT):\vT\vF=\vnull${\color{black}, where $\vT=\vC'(\vC\vC')^+\vC$ for an appropriate contrast matrix $\vC$}. These hypotheses are quite restrictive, in 
particular, no designs involving heteroscedastic variances are included in 
this set-up. Moreover, the test procedures are not consistent to detect arbitrary alternatives $H_1^F(\vT):\vT\vF\not = \vnull$. 
They are only consistent for alternatives of the form $H_1^p(\vT): \vT \vp \neq \vnull$
{\color{black}, where $\vp = (p_1, \ldots, p_d)'$ is the vector of the nonparametric effects 
$p_i$ defined in (2.3). As demonstrated in the analysis of the example in Section 8, these nonparametric effects describe a tendency to larger 
($p_i > \frac12$) or smaller ($p_i<\frac12$) values obtained from the distribution $F_i$ 
than randomly selected observations from the mean distribution $G=\frac1d \sumi d F_i$. 
Such an interpretation would not have been possible using only hypotheses formulated in terms of the distribution functions. 
Particularly in the data example it would have been difficult to demonstrate that the treatment is effective under both food conditions in a similar size. 
This is easily seen from Figure~4 in the supplementary material and enables an intuitive interpretation and visualization of the results for the practitioner.

Thus, it is reasonable to base nonparametric procedures on these effects $p_i$ since the alternative 
$H_1^p(\vT): \vT \vp \neq \vnull$ is the complement of the hypothesis 
$H_0^p(\vT): \vT \vp = \vnull$. 
These hypotheses are more general than the restrictive 
hypotheses formulated by the distribution functions.  In particular, hypotheses and effects in heteroscedastic designs which are commonly appearing in practice 
can be handled by this approach. Moreover, since the $p_i$ are fixed model quantities we can provide 
meaningful and intuitively interpretable confidence intervals for them. This would not have been possible using weighted relative effects 
(such as $r_i$ mentioned in Section~\ref{intro}) since they are no fixed model quantities that may lead to difficult interpretations as demonstrated in Table~\ref{relef} in the introduction.


Regarding technical considerations
}, the 
asymptotic \db\ of rank statistics based on such effects is quite difficult to 
handle since the asymptotic covariance matrix has a quite involved structure 
(see, e.g., Puri, 1964). All the more it appears difficult to derive estimators 
of the variances and covariances and to show their consistency. This problem is
overcome in the present approach by generating the vector $\vp = (p_1, \ldots,p_d)'$ of the relative effects $p_i$ as well as its estimator $\vwhp = (\hp_1, 
\ldots, \hp_d)'$ from the vector $\vw =(\vw_1',\vw_2', \dots, \vw_d')'$ of 
pairwise relative effects $\vw_i = (w_{1 i}, \ldots, w_{d i})' = \int \vF 
dF_i $ by the simple matrix multiplication in (\ref{vecrep}). By an in-depth study of the properties of $\vwhp$, test procedures for general null 
hypotheses $H_0^p(\vT): \vT \vp = \vnull$ {\color{black} in factorial designs and general split-plot layouts} have been developed.

\noindent Thus, the current gap between $H_0^F(\vT):\vT\vF = \vnull$ and the set of alternatives for which these tests are consistent has been closed by providing procedures for testing 
$H_0^p(\vT):\vT\vp = \vnull$. From these tests the ANOVA-type $p$-test 
$\varphi_N$ turned out to possess the best finite sample properties. 
{\color{black} We note, that} the Brunner-Munzel test (2000) for the nonparametric Behrens-Fisher problem is a special case of {\color{black} $\varphi_N$ if $d=2$}.





\section*{Acknowledgements}
{\color{black} The authors are grateful to two expert referees, the Associate Editor and the Editor 
for their helpful comments which led to a considerable improvement of the original version of the
paper.}\\ The work of M. Pauly was supported by the German Research Foundation 
project DFG-PA 2409/3-1.

\bcen
{\sc \large REFERENCES}
\ecen

\bdes\itemsep.5pt

{\footnotesize

\item {\sc Acion, L., Peterson, J. J., Temple, S., and Arndt, S.} (2006). 
Probabilistic index: An intuitive non-parametric approach to measuring the size 
of treatment effects. {\em Statistics in Medicine} {\bf 25}, 591--602.

\item {\sc Akritas, M.~G.} (1990). The rank transform method in some
two-factor designs. {\em J.\ Amer.\  Statist.\ Assoc.} {\bf 85}, 73--78.

\item {\sc Akritas, M.~G., and \ Arnold, S.~F.} (1994). Fully
nonparametric hypotheses for factorial designs I:  Multivariate repeated
measures designs. {\em J.\  Amer.\  Statist.\  Assoc.} {\bf  89},
336--343.

\item {\sc Akritas, M.~G., Arnold, S.~F., and Brunner, E.} (1997).
Nonparametric hypotheses and rank statistics for unbalanced factorial
designs. {\em J.\  Amer.\ Statist.\ Assoc.} {\bf 92}, 258--265.

{\color{black}
\item {\sc Akritas, M.~G. and  Brunner, E.} (1997). A unified approach to ranks 
tests in mixed models. {\em J.\  Statist.\ Plann.\ Inference} {\bf 61}, 249--277.
}

\item {\sc Akritas, M.~G.} (2011). Nonparametric Models for ANOVA and ANCOVA 
Designs. In {\em International Encyclopedia of Statistical Science}, Springer, 
964--968.


\item {\sc Bamber, D.} (1975). The area above the ordinal dominance graph and 
the area below the receiver operating characteristic graph. {\em Journal of 
Mathematical Psychology} {\bf 12}, 387--415.


\item {\sc Box, G.~E.~P.} (1954).
Some theorems on quadratic forms applied in the study of analysis of
variance problems, I. Effect of inequality of variance in the one-way
classification. {\em Ann.\  Math.\  Statist.} {\bf 25}, 290--302.

\item {\sc Boos, D. D., and Brownie, C.} (1992). A rank-based mixed model 
approach to multisite clinical trials. {\em Biometrics}, {\bf 44} 61--72.

\item {\sc Brown, B.M., and Hettmansperger, T.P.} (2002). Kruskal-Wallis, 
multiple comparisons and Efron dice.  {\em Australian \& New Zealand Journal of 
Statistics}, {\bf 44}, 427--438.

{\color{black}
\item {\sc Brumback, L.C., Pepe, M.S., and Alonzo, T.A.} (2006). Using the ROC 
curve for gauging treatment effect in clinical trials. {\em Statistics in 
Medicine} {\bf 25}, 575--590.
}

\item {\sc Brunner, E., Dette, H., and Munk, A.} (1997). Box-Type 
Approximations in Nonparametric Factorial Designs. {\em Journal of the American 
Statistical Association}, {\bf 92}, 1494--1502.

{\color{black}
\item {\sc Brunner, E., Domhof S.} and {\sc Langer, F.} (2002). {\em Nonparametric 
Analysis of Longitudinal Data in Factorial Designs}. Wiley, New York.  
}

{\color{black}
\item {\sc Brunner, E., Munzel, U. and Puri, M.~L.} (1999). Rank-Score Tests in Factorial Designs with Repeated Measures. {\em J. Mult. Analysis} {\bf 70}, 286--317. 

\item {\sc Brunner, E., Konietschke, F., Pauly, M. and Puri, M.L.} (2016). 
Supplementary Material to the paper Rank-Based Procedures in Factorial Designs: Hypotheses about Nonparametric Treatment Effects.
}

{\color{black}
\item {\sc Brunner, E. and Langer, F.} (2000). Nonparametric Analysis of 
Ordered Categorical Data in Designs with Longitudinal Observations and Small 
Sample Sizes. {\em Biometrical Journal} {\bf 42}, 663-675.
}

\item {\sc Brunner, E., and Munzel, U.} (2000). The Nonparametric Behrens-Fisher
Problem: Asymptotic Theory and a Small-Sample Approximation. {\em Biometrical
Journal} {\bf 42}, 17--25.

\item {\sc Brunner, E., and Munzel, U.} (2013). {\em Nichtparametrische 
Datenanalyse}. Springer, Heidelberg.

\item {\sc Brunner, E., and Puri, M.~L.} (2001). Nonparametric methods in 
factorial designs. {\em Statistical Papers} {\bf 42}, 1--52.

\item {\sc Brunner, E., and Zapf, A.} (2013). Nonparametric ROC Analysis for 
Diagnostic Trials, in {\em Handbook of Methods and Applications of Statistics 
in Clinical Trials}, Volume 2: Planning, Analysis, and Inferential Methods (N. 
Balakrishnan, Ed.), Wiley, 471--483.

{\color{black}

\item{\sc De Neve, Thas, O., Ottoy, J. P.,  and Clement, L.} (2013). An 
extension of the Wilcoxon-Mann-Whitney test for analyzing RT-qPCR data. {\em 
Statistical Applications in Genetics and Molecular Biology}, {\bf 12}, 333--346.

\item{\sc De Neve, J., Meys, J., Ottoy, J. P., Clement, L., and Thas, O.} (2014). 
The unified Wilcoxon-Mann-Whitney test for analyzing RT-qPCR data in R. 
{\em  Bioinformatics}, {\bf 30}, 2494--2495.
}

\item {\sc De Neve, J., and Thas, O.} (2015). A Regression Framework for Rank 
Tests Based on the Probabilistic Index Model. {\em J.\  Amer.\ Statist.\  
Assoc.}, {\bf 110}, 1276--1283.

{\color{black}
\item {\sc DiCiccio, C. J., and Romano, J. P.} (2015). Robust Permutation Tests for Correlation and Regression Coefficients. Technical Report No. 2015-15, Stanford University.
}
\item {\sc Domhof, S.} (2001). Nichtparametrische relative Effekte. Ph.D. 
Thesis, University of G\"ottingen. 

\item {\sc Fan, C., and Zhang, D.} (2014). Wald-type rank tests: A GEE 
approach. {\em Comp. Stat. Data Ana.,} {\bf 74}, 1--16.

{\color{black}

\item {\sc Fan, C., and Zhang, D.} (2015). On Power and Sample Size of the 
ANOVA-type Rank Test. {\em Communications in Statistics: Simulation and 
Computation}, to appear.

\item  {\sc Fischer, D., Oja, H., Schleutker, J., Sen, P. K., and Wahlfors, 
T.} (2014). {\em Generalized Mann-Whitney Type Tests for Microarray 
Experiments.} {\em Scandinavian Journal of Statistics}, {\bf 41}, 672--692.

\item {\sc Fischer, D., and Oja, H.} (2015). Mann-Whitney type tests for 
microarray experiments: the R package gMWT. {\em Journal of Statistical 
Software}, {\bf 65}, 1-19.}

\item {\sc Gao, X., and Alvo, M.} (2005). 
A unified nonparametric approach for unbalanced factorial designs. {\em J. 
Amer. Statist. Assoc.} {\bf 100}, 926--941.

\item {\sc Gao, X., and Alvo, M.} (2008). 
Nonparametric multiple comparison procedures for unbalanced two-way layouts.
{\em Journal of Statistical Planning and Inference} {\bf 138}, 3674--3686.

\item {\sc Gao, X., Alvo, M., Chen, J., and Li, G.} (2008). 
Nonparametric multiple comparison procedures for unbalanced one-way factorial
designs. {\em Journal of Statistical Planning and Inference} {\bf 138}, 
2574--2591.

\item {\sc Gardner, M.} (1970). The paradox of the nontransitive dice and the 
elusive principle of indifference. {\em Scientific American} {\bf 223}, 
110--114.

\item {\sc Hettmansperger, T.~P., and  Norton, R.~M.} (1987). Tests for
patterned alternatives in $k$--sample problems. {\em J.\  Amer.\
Statist.\  Assoc.} {\bf 82}, 292--299.

\item {\sc Janssen, A., and Pauls, T.} (2003). How do Bootstrap and permutation 
tests work$?$ {\em Annals of  Statistics} \textbf{31}, 768--806.
                     
\item {\sc Kaufmann, J, Werner, C., and Brunner, E.} (2005). 
Nonparametric methods for analysing the accuracy of diagnostic tests with 
multiple readers. {\em Statistical Methods in Medical Research} {\bf 14}, 
129--146.
                                           
\item {\sc Kieser, M., Friede, T., and Gondan, M.} (2013). Assessment of 
statistical significance and clinical relevance. {\em Statistics in Medicine} 
{\bf 32}, 1707--1719.

{\color{black}
\item {\sc Konietschke, F., Bathke, A. C., Hothorn, L. A., and Brunner, E.} (2010). 
Testing and estimation of purely nonparametric effects in repeated measures designs. 
{\em Computational Statistics and Data Analysis}, {\bf 54}, 1895--1905.
}

\item {\sc Konietschke, F., Hothorn, L. A., and Brunner, E.} (2012). 
{ Rank-based multiple test procedures and simultaneous confidence intervals.} 
{\em Electronic Journal of Statistics}, {\bf  6}, 738--759.


\item {\sc Kruskal, W. H., and Wallis, W. A.} (1952). Use of ranks in 
one-criterion variance analysis. {\em J.\  Amer.\ Statist.\  Assoc.}, {\bf 47}, 
583--621.


\item {\sc Lange, K.} (2008). Nichtparametrische Modelle f\"ur faktorielle 
Diagnosestudien, diploma thesis, University of G\"ottingen.


\item {\sc Mann, H. B., and Whitney, D. R.} (1947). On a test of whether one of 
two random variables is stochastically larger than the other. {\em Ann. Math. 
Stat.}, 50--60.

\item {\sc Mathai, A.M., and Provost, S.B.} (1992). {\it Quadratic Forms in 
Random Variables}. Marcel Dekker Inc., New York.

\item {\sc Munzel, U.} (1999). Linear rank score statistics when ties are 
present. {\em Stat. Prob. Lett.}, {\bf 41}, 389--395.




\item {\sc Pauly, M., Brunner, E., and Konietschke, F.} (2015a). Asymptotic 
permutation tests in general factorial designs. {\em Journal of the Royal 
Statistical Society: Series B} {\bf 77}, 461--473.

\item {\sc Pauly, M., Ellenberger, D., and Brunner, E.} (2015b). Analysis of 
High-Dimensional One Group Repeated Measures Designs. {\em Statistics: A 
Journal of Theoretical and Applied Statistics}, DOI: 
10.1080/02331888.2015.1050022.

\item {\sc Puri, M. L.} (1964). Asymptotic efficiency of a class of c-sample 
tests. {\em Ann. Math. Stat.}, 102--121.


{\color{black}
\item {\sc Rust, S.~W. and Fligner, M.~A.} (1984). A modification of the 
Kruskal-Wallis statistic for the generalized Behrens-Fisher problem. {\em 
Communications in Statistics - Theory and Methods} {\bf 13}, 2007--2013.
}
\item {\sc Ruymgaart, F.H.} (1980). A unified approach to the asymptotic 
distribution theory of certain midrank statistics. In: {\em Statistique
non Parametrique Asymptotique}, 1--18, J.P. Raoult (Ed.), Lecture Notes
on Mathematics, No. 821, Springer.

\item {\sc Thas, O., De Neve, J. D., Clement, L., and Ottoy, J. P.} (2012). 
Probabilistic index models. {\em Journal of the Royal Statistical Society: 
Series B}, {\bf 74}, 623--671.

\item {\sc Vallejo, G., Fern\'andez, M.P., and Livacic-Rojas, P. E.} (2010). 
Analysis of unbalanced factorial designs with heteroscedastic data. {\em 
Journal of Statistical Computation and Simulation}, {\bf 80}, 75--88.

{\color{black}
\item {\sc Vermeulen, K., Thas, O., and Vansteelandt, S.} (2015). Increasing 
the power of the Mann-Whitney test in randomized experiments through flexible
covariate adjustment. {\em Statistics in Medicine}, {\bf 34}, 1012--1030.

\item {\sc Zapf, A., Hoyer, A., Kramer, K., and Kuss , O.} (2015) Nonparametric 
meta-analysis for diagnostic accuracy studies. {\em Statistics in Medicine} 
{\bf 34}, 3831--3841.
}

}
\edes

\section{Appendix} \label{app}

{\bf Derivation of the covariances \eqref{sill}-\eqref{siill}.}
For ease of notation we will utilize the notation
$$\xi_{\ell i k} = F_\ell(X_{i k}) - w_{\ell i} \text{ and } \xi_{\ell i} = \xi_{\ell i 1} = F_\ell(X_{i 1}) - w_{\ell i}$$
and note that $E(\xi_{\ell i})=0$ and that $E(\xi_{\ell i}\xi_{\ell' i'})=0=$ whenever $i\neq i'$ due to independence. 
From this it follows in case of $i=i'\neq \ell=\ell':$ 
\bqa
s_i(l,l) &=& Var(\sqrt{N} Z_{\ell i}) = N Var(\frac{1}{n_i}\sum_{k=1}^{n_i} \xi_{\ell i k} - \frac{1}{n_\ell}\sum_{j=1}^{n_\ell} \xi_{i \ell j})\\ 
&=& \frac{N}{n_i} Var(\xi_{li}) + \frac{N}{n_\ell} Var(\xi_{il}) = \tau_i(\ell,\ell) + \tau_\ell(i,i)
\eqa
due to independence. Moreover, in case of $i=i'\neq \ell\neq\ell'\neq i$ we calculate

\bqa
s_i(l,l') 
&=& \Cov(\sqrt{N} Z_{\ell i}, \sqrt{N} Z_{\ell'i})\\
&=& N E[
  (\frac{1}{n_i}\sum_{k=1}^{n_i} \xi_{\ell ik} - \frac{1}{n_\ell}\sum_{k=1}^{n_\ell} \xi_{i \ell k})
  (\frac{1}{n_i}\sum_{j=1}^{n_i} \xi_{\ell' ij} - \frac{1}{n_{\ell'}}\sum_{j=1}^{n_{\ell'}} \xi_{i \ell' j})]\\
&=& \frac{N}{n_i^2} \sum_{k,j=1}^{n_i} E(\xi_{\ell ik} \xi_{\ell' ij}) = \frac{N}{n_i^2} \sum_{k=j=1}^{n_i} E(\xi_{\ell ik} \xi_{\ell' ik})\\
&=& \frac{N}{n_i} E(\xi_{\ell i} \xi_{\ell' i}) = \tau_i(\ell,\ell').
\eqa
In all other cases similar independence considerations show that $s_i(l,l') =0$. 
Concerning $s_{ii'}(l,l')$ we can proceed similarly by expanding
\bqan
s_{ii'}(l,l') 
&=& N E[
  (\frac{1}{n_i}\sum_{k=1}^{n_i} \xi_{\ell ik} - \frac{1}{n_\ell}\sum_{k=1}^{n_\ell} \xi_{i \ell k})
  (\frac{1}{n_{i'}}\sum_{j=1}^{n_{i'}} \xi_{\ell' i'j} - \frac{1}{n_{\ell'}}\sum_{j=1}^{n_{\ell'}} \xi_{i' \ell' j})].
\eqan
In case $i'\neq i=\ell'\neq\ell\neq i'$ this simplifies to 
\bqa
-\frac{N}{n_i^2} \sum_{k=1}^{n_i} E(\xi_{\ell ik} \xi_{i'i k}) = 
-\frac{N}{n_i} E(\xi_{\ell i} \xi_{i'i }) = -\tau_i(\ell, i')
\eqa
and the other cases are all analogue.\hfill\qed

{\bf Proof of Theorem~\ref{asFN}.}
Since $\sqrt{N} (\vwhp - \vp)$ is asymptotically normal with mean zero and covariance matrix $\vV$ the result follows from the continuous mapping theorem together with well-known results on quadratic forms, 
see e.g. Mathai and Provost (1992, p.29-36), where we implicitly utilized that $\vT = \vT' = \vT^2$.
\hfill\qed

{\bf Proof of Theorem~\ref{astests}.}\\
(a) Since $\vwhV$ is a consistent estimator for $\vV$ it holds that the difference between $Q$ and $\widehat{Q}$ converges to zero in probability, i.e. $\widehat{Q}-Q=o_P(1)$. 
Thus, due to continuity of the limit distribution,  $\hc(\alpha)$ converges in probability to the corresponding $(1-\alpha)$-quantile of $\mathcal{L}(Q)$, the distribution of $Q$. 
The result now follows from an application of Lemma~1 in Janssen and Pauls (2003).\\
(b) Fix $\vC\vp\neq 0$ and expand the enumerator of the test statistic as 
\bqa
\tr(\vT \vwhV_N) {\color{black}Q_N}(\vC) &=& N (\vwhp-\vp+\vp)' \vT (\vwhp-\vp+\vp) \\
&=& {N} \vp' \vT \vp  - 2 {N} (\vwhp-\vp)' \vT \vp + {N} (\vwhp-\vp)' \vT (\vwhp-\vp)\\
&=& N \left(\vp' \vT \vp + O_P(N^{-1/2}) + O_P(N^{-1})\right),
\eqa 
where the last equality follows from the asymptotic considerations in Section~\ref{asy}. Since $\vp' \vT \vp>0$ and $\tr(\vT \vwhV_N) \to \tr(\vT \vV)\neq 0$ in probability, it follows from Slutzky's Lemma that ${\color{black}Q_N}(\vC)\to +\infty$ in probability. 
This proves consistency of $\widehat{\varphi}_N = \mathbf{1}\{ {\color{black}Q_N}(\vC) > \hc(\alpha)\}$ since the distribution of $\widehat{Q}$ remains non-degenerated, see (a) above. 
Concerning $\widetilde{\varphi}_N = \mathbf{1}\{{\color{black}Q_N}(\vC) > \widehat{f}^{-1}\chi^2_{\widehat{f},1-\alpha}\}$ consistency follows from $\widehat{f}\geq 1$ (by Cauchy-Schwarz). 
Finally, the prove for $\varphi_N = \mathbf{1}\{{\color{black}Q_N}(\vC) > F_{1-\alpha}(\whf, \whf_1)\}$ follows similarly by noting that $\whf_1\to \infty$ in probability and thus $\whf^{-1}\chi^2_{\whf} \to 1$ in probability, see e.g. the Proof of Theorem 3.1.(a) 
in Pauly et al. (2015b).

\hfill\qed


\begin{center}
\textbf{\huge Supplementary Material}
\end{center}

\section{Extensions to General Repeated Measures Designs} \label{erm}

Some of the results from Section~5 are copied here for the readers convenience.

Let us consider a general nonparametric factorial repeated measures designs given by independent random vectors
\bqan \label{model rm}
\vX_{ik} = (X_{i\ell k})_{\ell=1}^t = (X_{i1k},\dots,X_{itk})', \quad \ig d; \ \kg n_i; \ \ell=1,\dots,t
\eqan
representing the  $t\in\natnu$ repeated measurements on subject $k$ in group $i$. 
As in the paper a factorial structure on the groups (whole-plot / between-subjects factors) and 
repeated measures (sub-plot / within-subjects factors) can be included by splitting up the indices $i$ and $\ell$, respectively. 
Also in this setting we can define adequate model parameters on the marginals of $X_{i\ell 1}\sim F_{i\ell}$. 
In particular, these are given by the relative effect of the distribution of group $i$ at time $\ell$ with respect 
to the unweighted pooled distribution function $G = \frac{1}{dt}\sum_{i=1}^d\sum_{\ell=1}^t F_{i\ell}$
\bqan
\label{releff rm}
p_{i\ell} &=& \int G dF_{i\ell} \quad \ig d; \  \ell=1,\dots,t.
\eqan
It can again be written as the mean $p_{i\ell} = \overline{w}_{\cdot\cdot i\ell}$ of the relative marginal effects 
 $w_{rsi\ell} = \int F_{rs} d F_{i\ell}, 1\leq i,r\leq d, 1\leq s,\ell\leq t,$ Collecting all $p_{i\ell}$ in a vector 
 $\vp =(p_{11},p_{12},\dots,p_{dt})'$ our linear hypotheses of interest can be written as $H_0^p:\vC\vp = \vnull$ for an adequate hypothesis matrices $\vC$.

For testing $H_0^p$ estimates for the effects $p_{i\ell}$ are obtained by substituting 
the distribution functions $F_{i\ell}(x)$ in \eqref{releff rm} with their empirical counterparts 
$\hF_{i\ell}(x) = \frac1{n_i} \sumk {n_i} c(x-X_{i\ell k})$ yielding 
\bqan \label{releffest rm}
 \hp_{i\ell} &=& \int \hG_t d \hF_{i\ell}.
\eqan
Thus, an estimator for the vector $\vp$ is given by $\vwhp =(\hp_{11},\hp_{12},\dots,\hp_{dt})'$ and its asymptotic behaviour can be studied similar to the univariate case. 
In particular, defining the vector
$$
\vw_{rs} = (w_{rs11},w_{rs12},\ldots,w_{rsdt})'
$$
and the matrix
$$
\vW = (\vw_{11}\vdots\vw_{12}\vdots\ldots\vdots\vw_{dt}) \in \renu^{dt\times dt}
$$
and denoting their empirical counterparts as $\vwhw_{rs}$ and $\vwhW$, respectively, it holds that
\bqan
\vp = \vE_{dt} \text{vec}(\vW) \quad \text{and}\quad \vwhp = \vE_{dt} \text{vec}(\vwhW).
\eqan
Here $\text{vec}$ denotes the usual matrix operator which stacks the columns of a matrix on top of each other and the matrix $\vE_{dt}$ is given by
$$
\vE_{dt} = \frac{1}{dt}\veins_{dt}'\otimes \vI_{dt}.
$$
Thus, by the asymptotic equivalence theorem, the random vector $\sqrt{N} (\vwhp - \vp)$ has the same asymptotic distribution as
$$
\sqrt{N}\vE_{dt} \vZ.
$$
Here $\vZ=(\vZ_{11}',\vZ_{12}',\ldots,\vZ_{dt}')'$ with $\vZ_{i\ell} = (Z_{11i\ell},Z_{12i\ell},\ldots,Z_{dti\ell})'$ and
$$
Z_{rsi\ell} = \frac1{n_i} \sumk {n_i}[F_{rs}(X_{i\ell k}) - w_{rsi\ell}] - 
\frac1{n_r} \sumk {n_r} [F_{i\ell}(X_{rsk}) - w_{i\ell rs}]
$$
denote sums of independent random variables. From this expression the following central limit theorem follows.

\bsa
 Let $\vV_N=\vE_{dt}\cov(\sqrt{N}\vZ)\vE_{dt}$. Then $\sqrt{N}(\vwhp - \vp)$ is asymptotically multivariate normally distributed with expectaion $\vnull$ and covariance matrix $\vV_N = \vE_{dt}\cov(\sqrt{N}\vZ)\vE_{dt}'$.
\esa

{\it Proof.} 

To apply the Cramer-Wold device let $\vk = (k_{11},\ldots, k_{dt})'$ denote an arbitrary vector of constants with $||\vk|| = 1$. 
It follows from the asymptotic equivalence result stated above that $\sqrt{N}\vk'(\vwhp - \vp)$ is asymptotically equivalent to
\bqa
&&\hspace{-.2cm}\sqrt{N}\sum_{i=1}^d\sum_{\ell=1}^t\left( \frac{1}{n_i} \sum_{k=1}^{n_i} k_{i\ell}(G(X_{i\ell k}) - F_{i\ell}(X_{i\ell k}))
-\frac{1}{dt} \underset{(r,s)\neq (i,\ell)}{\sum_{r=1}^{d}\sum_{s=1}^{t}}\frac{1}{n_r} \sum_{j=1}^{n_r}k_{i\ell} F_{i\ell}(X_{rsj}) + k_{i\ell}(1-2p_{i\ell})
\right)\\
&=&\sqrt{N}\sum_{i=1}^d\sum_{\ell=1}^t\left( \frac{1}{n_i} \sum_{k=1}^{n_i} k_{i\ell}(G(X_{i\ell k}) - F_{i\ell}(X_{i\ell k}))
-\frac{1}{n_i} \sum_{k=1}^{n_i} \frac{1}{dt} \underset{(r,s)\neq (i,\ell)}{\sum_{r=1}^{d}\sum_{s=1}^{t}}k_{rs} F_{rs}(X_{i\ell k}) + k_{i\ell}(1-2p_{i\ell})
\right)\\
&=&\sum_{i=1}^d\frac{\sqrt{N}}{n_i} \sum_{k=1}^{n_i} \tilde{Z}_{i k} ,
\eqa
where 
$$
\tilde{Z}_{i k} = \sum_{\ell=1}^t \left( k_{i\ell}(G(X_{i\ell k}) - F_{i\ell}(X_{i\ell k}))
- \frac{1}{dt} \underset{(r,s)\neq (i,\ell)}{\sum_{r=1}^{d}\sum_{s=1}^{t}}k_{rs} F_{rs}(X_{i\ell k}) + k_{i\ell}(1-2p_{i\ell})\right)
$$
are independent random variables with expectation zero. Since these random variables are uniformly bounded it follows from the 
Lindeberg-Feller central limit theorem and the Cramer-Wold device that $\sqrt{N}(\vwhp - \vp)$ is asymptotically multivariate normally 
distributed with expectation $\vnull$ and covariance matrix $\vV_N = \vE_{dt}\cov(\sqrt{N}\vZ)\vE_{dt}'$.

Since the involved covariance matrix $\vSigma = (\Sigma_{rsi\ell})\equiv \cov(\sqrt{N}\vZ)$ is unknown we have to estimate it. 
Therefore, we first analyze its explicit form and proceed as in 
Placzek (2013).
First, consider the case $i=r$ and $s=\ell$ and set
$$
\Sigma_{rsrs} = \left(N \cov(Z_{pqrs},Z_{p'q'rs}) \right)_{1\leq p,p'\leq d, 1\leq q,q'\leq t}
\equiv \left(\sigma_{rs}(p,q,p',q')\right)_{1\leq p,p'\leq d, 1\leq q,q'\leq t}.
$$
Since $Z_{rsrs}=0$, $Z_{rsi\ell}=-Z_{i\ell rs}$ and $X_{i\ell k}$ is independent from $X_{i'\ell' k'}$ for all $i\neq i'$ or $k\neq k'$ it 
follows that $\sigma_{rs}(p,q,p',q')=$ 
\bqa
\left\{ \begin{array}{c}
                                 \tau_r^{(s,s)}(p,q,p',q')\\
                                 \tau_r^{(s,s)}(p,q,p,q')  +  \tau_p^{(q,q')}(r,s,r,s)\\
                                 \tau_r^{(s,s)}(r,q,p',q') - \tau_r^{(q,s)}(r,s,p',q')\\
                                 \tau_r^{(s,s)}(p,q,r,q')  -  \tau_r^{(s,q')}(p,q,r,s)\\
				 \tau_r^{(s,s)}(p,q,r,q')  -  \tau_r^{(s,q')}(r,q,r,s)- \tau_r^{(q,s)}(r,s,r,q')  +  \tau_r^{(q,q')}(r,s,r,s)\\
				 0
\end{array}
\right.
if 
\begin{array}{c} 
 r\neq p, p'\wedge p \neq p'\\
 r\neq p, p'\wedge p = p'\\
 r = p\wedge p' \neq p'\wedge q \neq s\\
 r = p\wedge p' \neq p\wedge q' \neq s\\
  r = p= p' \wedge q \neq  s \wedge q'\neq s\\
\text{else.}
  \end{array}
\eqa
Here 
\bqan\label{tau:1}
 \tau_r^{(s,\ell)}(p,q,p',q') &=& \frac{N}{n_r} 
 E \left[ \left(F_{pq}(X_{rs1}) - w_{pqrs} \right) 
\left(F_{p'q'}(X_{r\ell1}) - w_{p'q'r\ell} \right) \right].
\eqan

Now, consider the case $(r,s)\neq (i,\ell)$ and set
$$
\Sigma_{rsi\ell} = \left(N \cov(Z_{pqrs},Z_{p'q'i\ell}) \right)_{1\leq p,p'\leq d, 1\leq q,q'\leq t}
\equiv \left(\sigma_{rsi\ell }(p,q,p',q')\right)_{1\leq p,p'\leq d, 1\leq q,q'\leq t}.
$$
From similar considerations as above it follows for each entry that $\sigma_{rsi\ell}(p,q,p',q')=$ 
\bqa
\left\{ \begin{array}{cc} 
\tau_r^{(s,\ell)}(p,q,p',q') &  r=i \wedge p \not = i, p' \wedge r \not = p' \\
-\tau_r^{(s,q')}(p,q,i,\ell) & r = p' \wedge p \not = i, p' \wedge r \not = i \\
-\tau_p^{(q,q')}(r,s.i,\ell) & p=i \wedge r \not = i, p' \wedge p \not = p'\\
\tau_p^{(q,q')}(r,s,i,\ell) & p = p' \wedge r \not = i, p'\wedge p \not = i\\
\tau_r^{(s,\ell)}(p,q,r,q') - \tau_r^{(s,q')}(p,q,r,j)& r = i = p' \wedge p \not = i, p' \wedge q' \not = \ell\\
-\tau_p^{(q,j)}(r,s,p,q') + \tau_p^{(q,q')}(r,s,i,\ell)& p = i = p' \wedge r \not = i, p' \wedge q' \not = \ell\\
\tau_r^{(s,\ell)}(r,q,p',q') - \tau_r^{(q,\ell)}(r,s,p',q')& r=i=p \wedge p' \not = i, p \wedge q \not = s\\
-\tau_r^{(s,q')}(r,q,i,\ell) + \tau_r^{(q,q')}(r,s,i,\ell)& p = r = p' \wedge i \not = r,p \wedge q \not = s\\
\tau_r^{(s,\ell)}(p,q,p,q') + \tau_p^{(q,q')}(r,s,r,j)& r = i \wedge p = p' \wedge r \not = p' \wedge p \not = i\\
-\tau_r^{(s,q')}(p,q,p,\ell) -\tau_p^{(q,\ell)}(r,s,r,q')& r = p'\wedge p = i \wedge r \not = i \wedge p \not = p'\\
\tau_r^{(s,\ell)}(r,q,r,q')-\tau_r^{(s,q')}(r,q,r,\ell)
-\tau_r^{(q,\ell)}(r,s,r,q')+\tau_r^{(q,q')}(r,s,r,\ell)& r = p = p' = i \wedge q \not = s \wedge q' \not =  \\
\text{}  & \ell\wedge q \not = \ell \wedge s \not = q' \wedge q \not = q'\\
0 & else
\end{array}
\right.
\eqa

Thus, for estimating the unknown covariance $\vV_N$ we only have to estimate the unknown quantities given in \eqref{tau:1}. Similar to the paper consistent 
estimaors $\whtau_r^{(s,\ell)}(p,q,p',q')$ are obtained by calculating the arithmetic means of the empirical counterparts of \eqref{tau:1}. This yields a consistent estimator $\vwhV_N$ of $\vV$ and 
an ANOVA-type-statistic for $H_0^p$ is given by 
\bqan \label{ATS}
 Q_N(\vC) \ = \ Q_N(\vT) &=&  \frac{N }{\tr(\vT \vwhV_N)} \vwhp' \vT \vwhp, 
\eqan
where again $\vT=\vC'(\vC\vC')^+\vC$ is the unique projection matrix on the column space of $\vC$, see e.g. Brunner et al. (1997) or Brunner and Puri (2001). 
As in Theorem~4.1 of the paper $Q_N(\vC)$ has, asymptotically under the null $H_0^p :\vT \vp=\vnull$, the same distribution as a weighted sum of independent $\chi_1^2$-distributed random variables. 
An {\it ANOVA-eigen-type-$p$-test} can be obtained by estimating the unknown weights using  the consistent matrix estimator $\vwhV_N$. 
The investigation of this approach will be part of future work together with a simultaneous inference procedure.

\section{Interpretation of the Nonparametric Effects} \label{int}

Here we outline an interpretation of the nonparametric effects by using a decomposition of the distribution functions as in Akritas and Arnold (1994). 
It is similar to the interpretation for the relative effects considered in the supporting information in de Neve and Thas (2015). 
For ease of convenience we only consider the situation of a crossed two-way layout. 
To this end, write
$$
 F_{ij}=G+A_i+B_j+(AB)_{ij} \qquad(i=1,\dots,a; j=1,\dots,b)
$$
for functions $G, A_i, B_j, (AB)_{ij}$ satisfying $\sum_{i=1}^a A_i= \sum_{j=1}^b B_j =0$, $\sum_{i=1}^a (AB)_{ij}=0$ for all 
$j=1, \ldots, b$ and $\sum_{j=1}^b (AB)_{ij}=0$ for all $i=1, \ldots, a$. This expression is related to the classical mean decomposition in linear models. 
In particular, we can write $G= \olF_{\cdot \cdot} = \frac{1}{ab}\sum_{i=1}^a\sum_{j=1}^b F_{ij}$, $A_i=\olF_{i\cdot}-G = \frac{1}{b}\sum_{j=1}^b F_{ij} - G$, 
$B_j=\olF_{\cdot j}-G = \frac{1}{a}\sum_{i=1}^a F_{ij} - G$ and $(AB)_{ij}=F_{ij}-\olF_{i\cdot}-\olF_{\cdot j}+G$, for $i=1, \dots, a, j=1,\dots,b$. 
Inserting the above decomposition into 
the nonparametric effects $p_{ij}$ now results in
\bqa
 p_{ij} &=& \int G d F_{ij} \ =\  \frac{1}{2} + \underbrace{\int G d A_i}_{\alpha_i} + \underbrace{\int G d B_j}_{\beta_j} + \underbrace{\int G d(AB)_{ij}}_{(\alpha\beta)_{ij}}.
\eqa 
Here the additive effects all fulfill the side conditions $\sum_{i=1}^a \alpha_i= \sum_{j=1}^b \beta_j =0$, $\sum_{i=1}^a (\alpha\beta)_{ij}=0$ for all 
$j=1, \ldots, b$ and $\sum_{j=1}^b (\alpha\beta)_{ij}=0$ for all $i=1, \ldots, a$ which they inherit from the corresponding functions. Thus, as in the supporting information 
in de Neve and Thas (2015), we can interpret the additive effect $\alpha_i$ as
$$
 \alpha_i = \int G dF_{i\cdot} -\frac{1}{2} = P(Z_G< Z_{ai}) + \frac{1}{2}P(Z_G = Z_{ai}) - \frac{1}{2},
$$
where $Z_G\sim G$ and $Z_{ai}\sim \olF_{i\cdot}$. Similar interpretations hold for $\beta_j$ (see the paper for details) and $(\alpha\beta)_{ij}$, respectively.

\section{More Details on the Analysis of the Data Example}

Some parts from Section~8 are copied here for the readers convenience.\\

In order to provide freely available software for data analysis and educational purposes we implemented 
an {\it R} software package called {\bf rankFD} for rank based analysis of factorial designs with independent observations. 
For a user-friendly implementation it is equipped with a graphical user interface. 
The package contains the ANOVA-type-p-test (who turned out to be the best in our simulation study) for making inference in one-, two- or arbitrary higher-way layouts 
as well specific nested designs. Furthermore, all test procedures for testing the hypothesis $H_0^F$ formulated in terms of the distribution functions are implemented. 
Besides of a descriptive overview it also provides $p$-values and confidence intervals for the main treatment effects along with plotting options. The \textit{R} package will be updated 
regularly. 
The {\it R}-package is freely available at CRAN. Here it has been exemplified for analysing the motivating data example described in the Introduction of the paper and in more detail below.

In a placebo-controlled trial, the effect of a drug on the immune system
was examined under consideration of stress (food deprivation) using 40 mice.
A main response variable was the number of leucocytes migrating into the 
peritoneum. Half of the mice received a diet low in protein, the other half 
received normal food. One day before opening the peritoneum, 20 mice in each 
group received an injection with the drug, while the other 20 received an equal 
amount placebo. Eight hours later, migration of leucocytes was stimulated by 
injecting glycogen into every mouse. Then, for the resulting four groups, the 
number of leucocytes (among other attributes) was determined for each mouse. 
Because of copy right and confidentiality reasons only a part of the data from 
the complete trial is given in Table~\ref{leukanz.dat}. We are grateful to Fa. 
Schaper \& Br\"ummer (Salzgitter) for making available these data from a common 
research project.

\begin{table}[H]
\caption{Number of Leukocytes [$10^6/ml$] for 40 mice. All combinations of the 
following two treatments were examined: normal diet vs. low protein diet and 
drug vs. placebo.} 
\label{leukanz.dat}
\bcen
{\small
\btb{crccrccrccrc} \hline
\mc{12}{c}{Number of Leukocytes [$10^6/ml$]} \\ \hline
\mc{6}{c}{Normal Food} & \mc{6}{c}{Reduced Food} \\ \hline
\mc{3}{c}{Placebo} & \mc{3}{c}{Drug} & \mc{3}{c}{Placebo} & \mc{3}{c}{Drug} \\
\hline \hline
\qquad & 7.5 & \qquad & \qquad & 15.9 & \qquad & \qquad & 7.5 & \qquad & \qquad & 5.7 
& \qquad \\
\qquad & 8.1 & \qquad & \qquad & 12.0 & \qquad & \qquad & 5.7 & \qquad & \qquad & 8.1
& \qquad \\
\qquad &  5.4 & \qquad & \qquad & 12.3 &\qquad & \qquad & 3.3 & \qquad & \qquad &  6.0  \\
\qquad &  6.0 & \qquad & \qquad & 44.4 &\qquad & \qquad & 3.9 & \qquad & \qquad &  6.0  \\
\qquad & 16.2 & \qquad & \qquad & 13.5 &\qquad & \qquad & 3.9 & \qquad & \qquad & 11.4  \\
\qquad &  7.8 & \qquad & \qquad & 19.8 &\qquad & \qquad & 6.6 & \qquad & \qquad &  5.1  \\
\qquad &  8.1 & \qquad & \qquad & 15.3 &\qquad & \qquad & 6.3 & \qquad & \qquad & 11.1  \\
\qquad &  5.7 & \qquad & \qquad & 32.7 &\qquad & \qquad & 3.3 & \qquad & \qquad & 12.9  \\
\qquad &  6.9 & \qquad & \qquad & 18.0 &\qquad & \qquad & 4.5 & \qquad & \qquad &  5.4  \\
\qquad &  5.1 & \qquad & \qquad & 15.0 &\qquad & \qquad & 4.2 & \qquad & \qquad &  8.4  \\ \hline
\etb
} 
\ecen
\end{table}

{\color{black}
Applying the {\it R}-package {\bf rankFD} to the above data set yields the following statistics and $p$-values for testing the main effects $A$ ({\it food
condition}) and $B$ ({\it treatment}) as well as the interaction $AB$ between
the food condition and the treatment shown in Table~\ref{Table Ex}. 
}

\begin{table}[H]
\caption{Analysis of the data example with the ANOVA-type$-p$-test $\varphi_N$
{\color{black}given in Theorem~4.2b)(3).} The value of the test statistic $Q_N(\vT)$ is compared with the quantile of an
$F$-distribution with estimated degrees of freedom $\whf_1$ and $\whf_2$.}
\label{Table Ex}
{\small
\bcen
\begin{tabular}{lrccr}
Factor & Statistic & $\whf_1$ & $\whf_2$ & $p$-value \\\hline
Food {\color{black}Condition}& 42.450 & 1 & 26.492 & $< 0.0001$ \\
Treatment & 33.191 & 1 & 26.492 & $< 0.0001$ \\
Interaction &  1.868 & 1 & 26.492 & $0.1832$ \\
\hline
\end{tabular}
\ecen
}

\end{table}

It {\color{black}appears} from Table~\ref{Table Ex} that both the factors \textit{Food} as
well as \textit{Treatment} have a significant impact on the numbers of
leucocytes at 5\% level. The data do not provide {\color{black}any} evidence {\color{black}for an interaction between the treatment and the food condition.

%
So far the state of the art nonparametric approach using ranks to test the null hypotheses of no treatment effects or no interaction 
would have been using the procedures based on the \dfs\
$F_{ij}(x)$, i.e. $H_0^F: \vT \vF = \vnull$, where $\vT$ denotes an appropriate contrast matrix (for details see Akritas et al., 1997). 
the hypothesis of no food effect would be written as $H_0^F(A): F_{11} + F_{12} - F_{21} - F_{22} \equiv 0.$
Here the index $i$ in $F_{ij}$ refers to the factor $A$ (food condition: $i = 1$, normal food; $i = 2$, reduced food) 
while the second index $j$ refers to the factor $B$ (treatment: $j = 1$, placebo; $j = 2$, drug). 
A rejection or acception of these hypotheses would help for a first intuition about underlying effects, however, the testing procedures 
would not help for deducing the same elaborated interpretations and conclusions as done with the unweighted relative effects $p_{ij}$ in Section~8 of the paper. 
The only possibility for more intuition would be to plot the empirical versions of the sums and differences of distribution functions defining the hypotheses. 
To demonstrate this we plot the so-called empirical interaction function $x\mapsto (\widehat{AB}_{11})(x) = \frac{1}{4}[\hF_{11}(x) - \hF_{12}(x) - \hF_{21}(x) + \hF_{22}(x)]$ and the 
empirical main effect functions $x\mapsto (\widehat{A}_{1})(x) = \frac{1}{4} [\hF_{11}(x) + \hF_{12}(x) - \hF_{21}(x) - \hF_{22}(x)]$ and 
$x\mapsto (\widehat{B}_{1})(x) = \frac{1}{4} [\hF_{11}(x) - \hF_{12}(x) + \hF_{21}(x) - \hF_{22}(x)]$ in Figures~\ref{plot: interact}-\ref{plot: mainb} below.

\begin{figure}
\bcen
\includegraphics[scale=0.4]{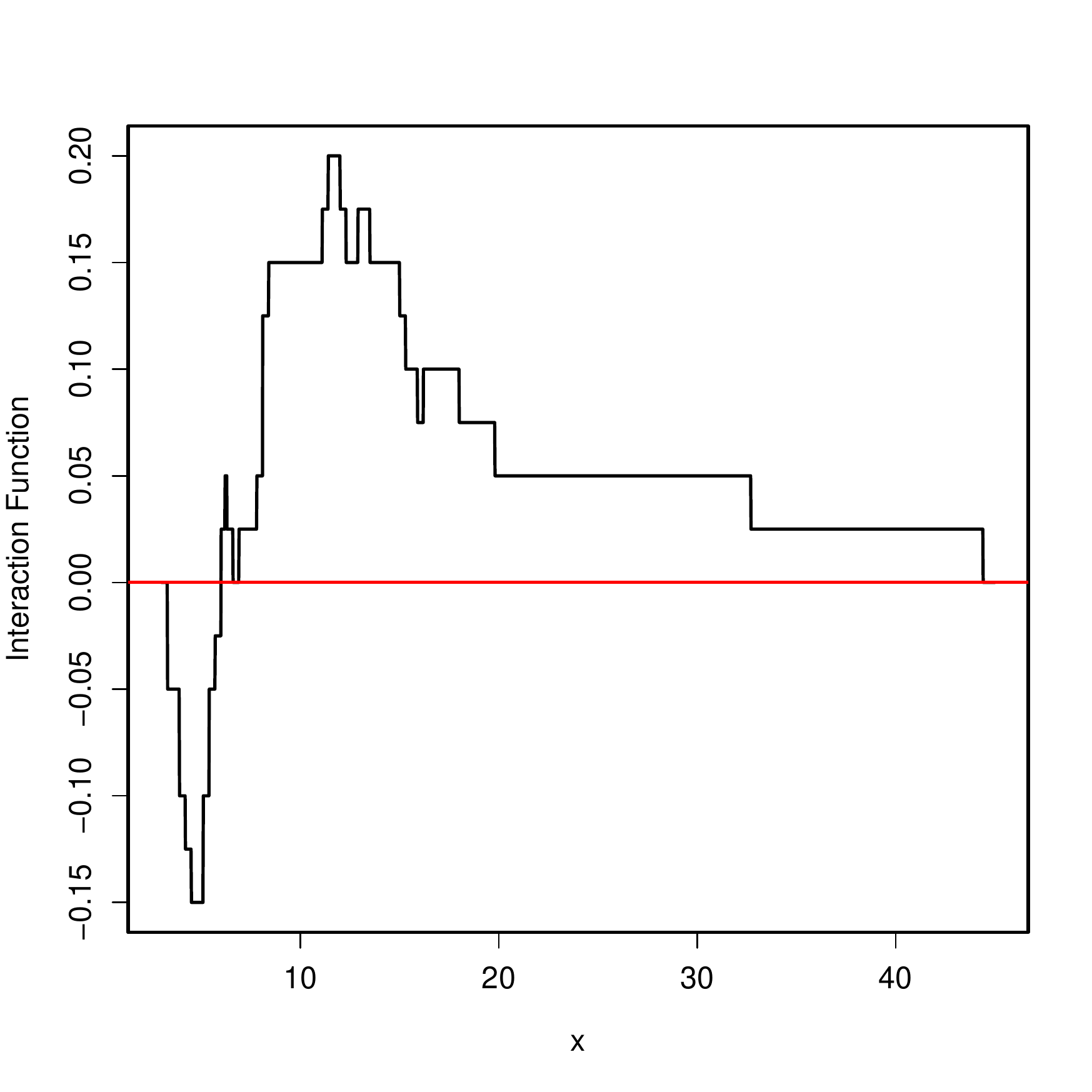}
\ecen
\caption{Plot of the empirical interaction function $(\widehat{AB}_{11}): x\mapsto \frac{1}{4}[\hF_{11}(x) - \hF_{12}(x) - \hF_{21}(x) + \hF_{22}(x)]$} \label{plot: interact}
\end{figure}

\begin{figure}
\bcen
\includegraphics[scale=0.4]{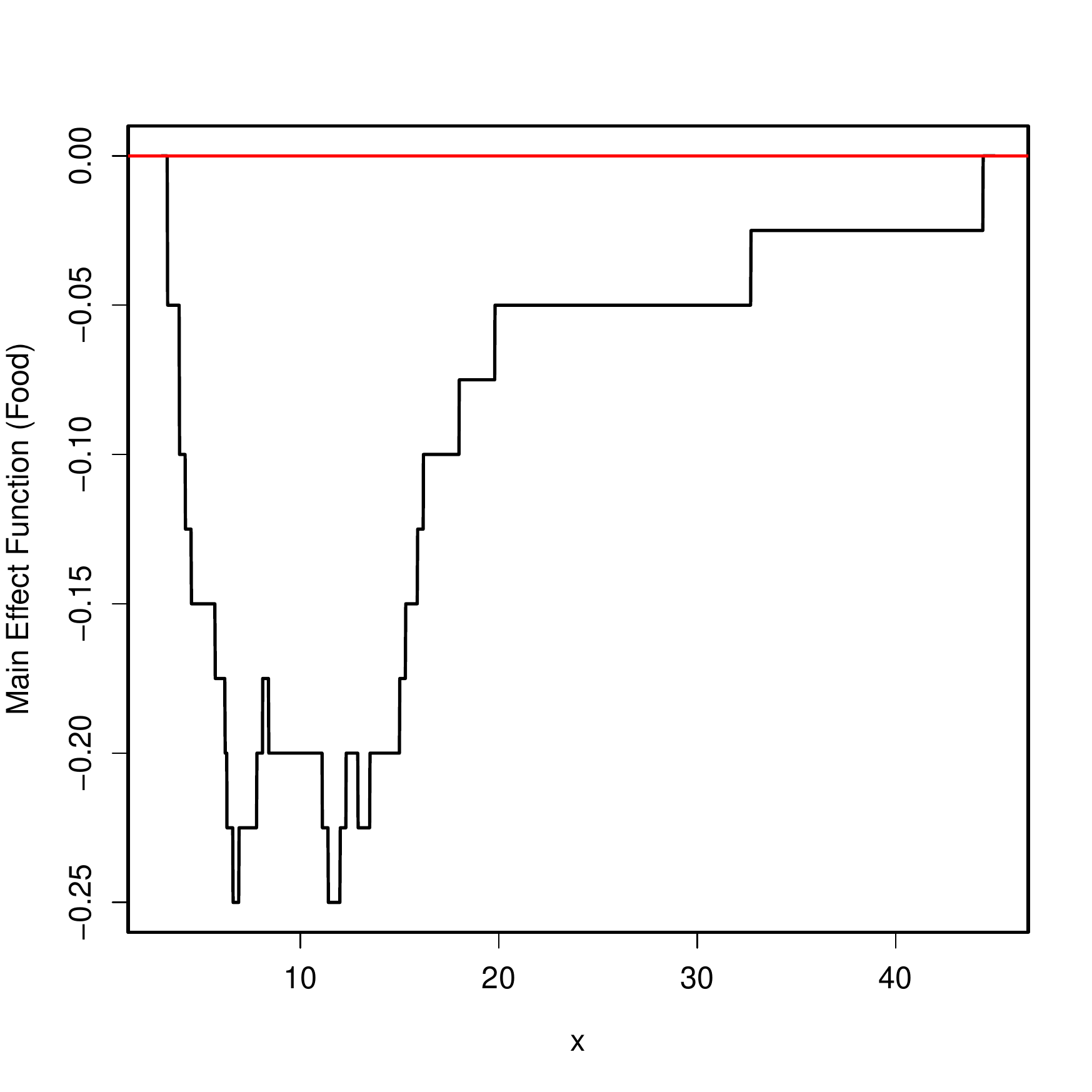}
\ecen
\caption{Plot of the empirical main effect function $(\widehat{A}_{1}):x\mapsto \frac{1}{4}[\hF_{11}(x) + \hF_{12}(x) - \hF_{21}(x) - \hF_{22}(x)]$} \label{plot: maina}
\end{figure}

\begin{figure}
\bcen
\includegraphics[scale=0.4]{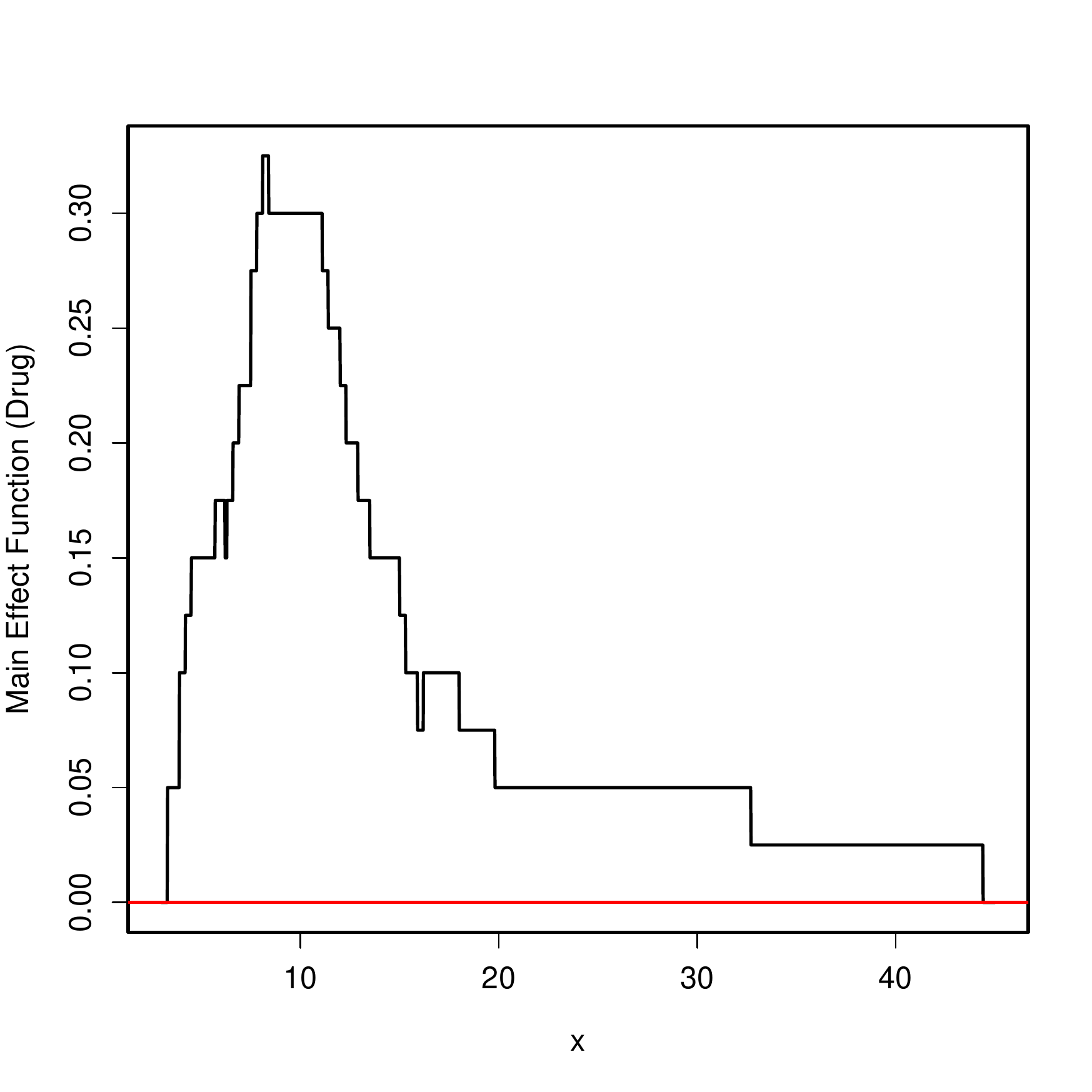}
\ecen
\caption{Plot of the empirical main effect function $(\widehat{B}_{1}):x\mapsto  \frac{1}{4}[\hF_{11}(x) - \hF_{12}(x) + \hF_{21}(x) - \hF_{22}(x)]$} \label{plot: mainb}
\end{figure}

From Figures~\ref{plot: maina} and \ref{plot: mainb} 
it is obvious that the main effect functions are different from the $0$-function.
But this is also true for the plot of the empirical interaction function in Figure~\ref{plot: interact}. 
No intuitive conclusion regarding an interaction can be drawn from this figure. 
This demonstrates the gap between the hypotheses of the procedures based on $H_0^F$ 
and the set of alternatives for which they are consistent. 
One must note that the above main and interaction effects defined by the \dfs\ are functional-valued quantities which are difficult to interpret.

This is different, however, for the nonparametric effects $p_{ij} = \int G dF_{ij}$ 
considered in the main paper. 
For these real-valued effect measures point estimators for each drug$\times$food combination can be computed. 
Also two-sided (range preserving) 95\%-confidence intervals for the $p_{ij}$ are computed where the logit
transformation $g(x) = \log(x / (1-x))$ has been used. The results are listed in
Table~\ref{TableDescriptive} and displayed in Figure~\ref{CI}. 

\begin{figure}
\bcen
\includegraphics[scale=0.4]{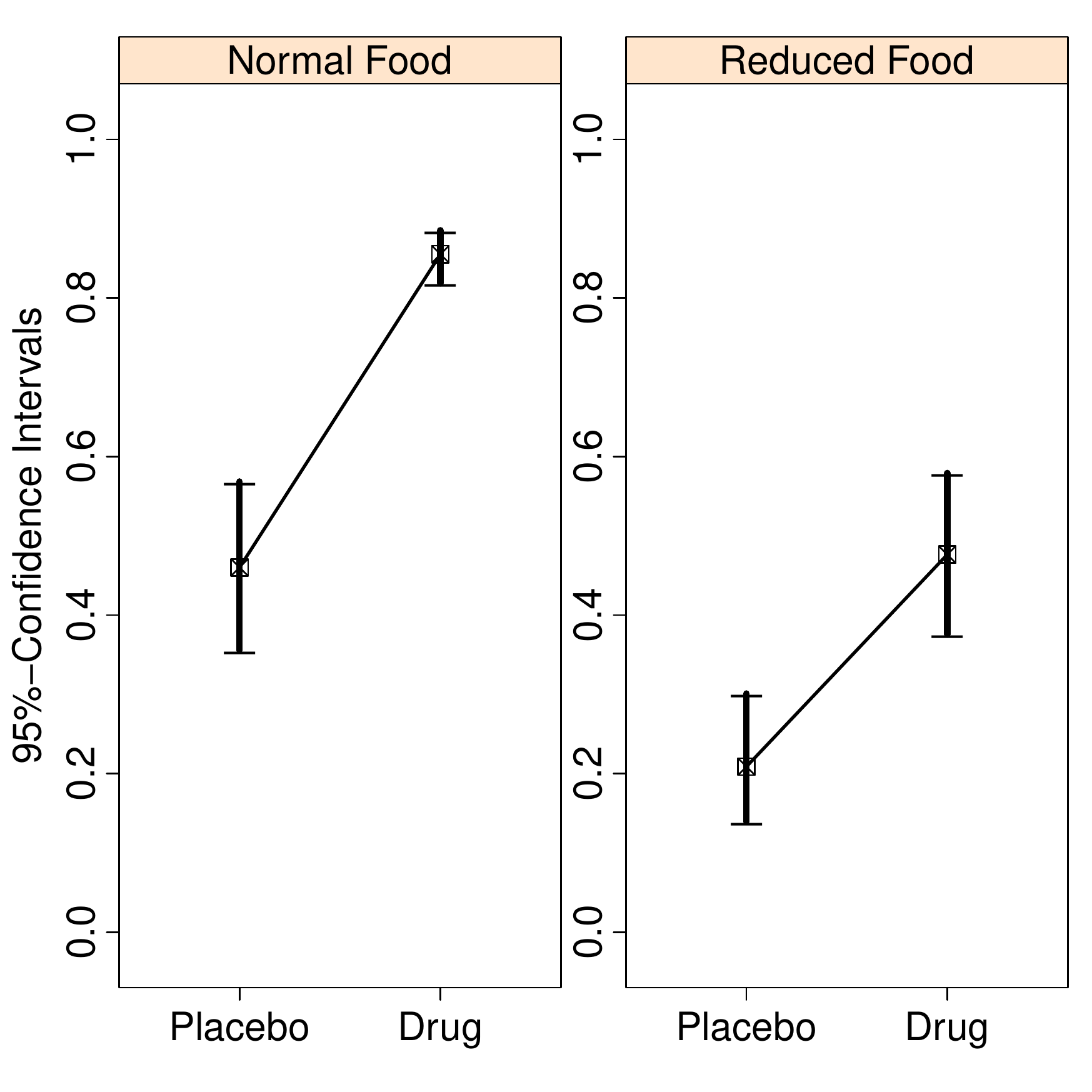}
\ecen
\caption{Plot of the $95\%$ confidence intervals for each drug$\times$ food combination.} \label{CI}
\end{figure}

\begin{table}[H]\color{black}
\caption{Estimates and 95\%-confidence intervals for the nonparametric treatment effects $p_{ij} = \int G dF_{ij}$ in the leucocytes trial. The index
$i$ refers to the food condition while the index $j$ refers to the treatment.
The range-preserving limit are obtained by the logit-transformation $g(x) =
\log(x / (1-x))$.}
\label{TableDescriptive}
\bcen
\begin{tabular}{llclll}\hline
\mc{2}{c}{Factor Level Combination} & Sample Size & \mc{1}{c}{Effect} &
\mc{2}{c}{95\%-Confidence Limits}\\ \hline
\mc{1}{c}{Food Condition} & \mc{1}{c}{Treatment} & $n_{ij}$ &
\mc{1}{c}{$\hp_{ij}$} & Lower & Upper \\ \hline
$i=1$ \ - \ Normal & $j=1$ \ - \ Placebo & 10 & 0.460 & 0.355 & 0.568 \\ 
$i=1$ \ - \ Normal & $j=2$ \ - \ Drug & 10 & 0.855 & 0.818 & 0.885   \\
$i=2$ \ - \ Reduced & $j=1$ \ - \ Placebo & 10 & 0.209 & 0.140 & 0.301 \\
$i=2$ \ - \ Reduced & $j=2$ \ - \ Drug & 10 &  0.476 & 0.375 & 0.579 \\ \hline
\end{tabular}
\ecen
\end{table}
\newpage

The effect $\hp_{21} = 0.209$ for the reduced food under placebo means that the
observations from $F_{21}$ tend to be smaller than those from the mean
distribution $G=\tfrac{1}{4}\sum_{i,j=1}^2 F_{ij}$, or more precisely, the probability that a randomly selected 
observation $Z$ from the mean distribution $G$ is smaller than a randomly selected
observation $X_{21}$ from $F_{21}$ equals $0.209$. Similarly, the effect $\hp_{12} =
0.855$ for the normal food under the drug means that the observations from
$F_{12}$ tend to be larger than those from the mean distribution $G$. 
We note that the confidence intervals for Placebo and Drug do not overlap within each food condition which may be interpreted that the drug is effective in both cases, as seen from Figure~\ref{CI}. 
Such an interpretation is difficult to conclude from plots of the empirical effect functions.

\section{More Simulation Results}

Again most parts from Section~7 are copied here for the readers convenience.\\

Here we investigate the small sample properties of the three statistical tests $\widehat{\varphi}_N$, $\widetilde{\varphi}_N$, and $\varphi_N$ based on the 
ANOVA-type statistic ${Q_N}(\vT)$ and given in Theorem~4.2.(b) within extensive simulation studies with regard to their
\bit
\item[(a)] maintenance of the preassigned type I error level ($\alpha = 5\%$) under the hypothesis $H_0^p(\vT): \vT \vp = \vnull$ and 
\item[(b)] their powers to detect specific alternatives.
\eit

All simulations were performed using {\it R} (version 2.15.0, R Development Core Team, 2010) with $nsim=10,000$ simulation runs for each setting. 
As in the main paper the distribution of $\whQ(\vT)$ was approximated using $n_{MC}=10,000$ Monte-Carlo runs, and the critical values 
were estimated from this distribution. Hereby, the eigenvalues of the matrix $\vT\vwhV_N$ were computed with the base {\it R}-function \textit{eigen}. 

In order to compare the newly developed methods with {\color{black} other procedures we first} 
restrict our considerations to the one-way layout (balanced and unbalanced) with $a=4$ 
independent treatment groups, and by using both symmetric and skewed distributions. 
{\color{black} In this set-up the above procedures test the null hypothesis $H_0^p: p_1=p_2=p_3=p_4$. 
As competitors the classical Kruskal-Wallis rank test and two Wald-type tests are considered: 
The test ${\color{black}\varrho_N} = \veins\{{\color{black}W_N}(\vC) > \chi^2_{1-\alpha;r({\color{black}\vwhM_N})}\}$ based on the WTS given in Section~4 of the paper 
and a related test in a Wald-type statistic for a probabilistic index model (PIM, Thas et al., 2012) using a sandwich-type covariance matrix estimator, say $\vwhS$, 
and weighted rank estimators for the PIM effects, say $\vwhalpha$, instead of $\vwhV_N$ and $\vwhp$, respectively, and a $\chi^2$-quantile with estimated degrees of freedom given by $r(\vC\vwhS\vC')$. 
The latter is motivated from the considerations in de Neve and Thas (2015) {\color{black}and denoted as DTS}. 
We note that it is a test for the related null hypothesis $H_0^\alpha: \alpha_1=\dots=\alpha_4$ formulated in terms of the weighted 
PIM effects $\alpha_i$ (see Equation (4) in de Neve and Thas, 2015, for its explicit definition) which is equal to $H_0^p$ in the balanced case. 
The ingredients of the test statistic were calculated as described in the supplementary material of de Neve and Thas (2015) with the {\it R} package {PIM} (Version 1.1.5.6). 
Moreover, note that the Kruskal-Wallis test has been developed for testing the more restrictive null hypothesis $H_0^F: F_1=F_2=\dots=F_a$ formulated in terms of the distribution functions.
}

Symmetrically distributed data was generated from the model 
\bqa X_{ik} = \mu_i  +\sigma_{i} \epsilon_{ik}, \quad i=1,\ldots,a; \sep k=1,\ldots,n_{i},
\eqa
where the random error terms
\bqa
\epsilon_{ik}= \frac{\wtep_{ik} - E(\wtep_{i1})}{\sqrt{\Var(\wtep_{i1})}}
\eqa 
were generated from different standardized symmetric distributions, i.e., the random variables $\wtep_{ik}$ were generated from 
standard normal {\color{black} or the}  double exponential distribution, respectively. 
 Skewed data was generated from log-normal-distributions by $X_{ik} = \exp(\eta_{ik})$, where $\eta_{ik}\sim N(0,\sigma_i^2)$ and possibly different variances $\sigma_i^2$. Note that the null hypothesis $H_0^p: \vP_a\vp=\vnull$ holds in both cases, because of the symmetry and the monotonicity of the exponential function.

A major assessment criterion for the accuracy of the methods is their behavior when different sample sizes and variances are combined, i.e. when increasing sample sizes are combined with increasing variances (\textit{positive pairing}) or with decreasing variances (\textit{negative pairing}) 
(see Pauly et al., 2015a). We consider balanced situations with sample size vector $\vn_1=(n_1,n_2,n_3,n_4)=(5, 5, 5, 5)$ and unbalanced situations with sample size vector $\vn_2=(n_1,n_2,n_3,n_4)=(10,20,30,40)$, respectively. The scaling vector $\vsigma=(\sigma_1,\sigma_2,\sigma_3,\sigma_4)$ was chosen from $(1,1,1,1), (1,\sqrt{2},2,\sqrt{5})$ or $(\sqrt{5},2,\sqrt{2},1)$, respectively. In order to investigate the behavior of the tests when the sample sizes increase, a constant $m \in \{5,10,20,25\}$ was added to each component of the vectors $\vn_1$ and $\vn_2$, i.e. $\vn_i + m\veins_4' = (n_1 + m, n_2 + m, n_3 + m, n_4 + m), i=1,2$. The different simulation settings are summarized in Table~\ref{SimSetting1}. 
\newpage

\begin{table}[h]
\caption{Simulated one-way layout with $a=4$ samples, where $m \in \{0, 5, 10, 20, 25\}$ and $\vn_1=(5, 5, 5, 5)$, and $\vn_2=(10,20,30,40)$. } \label{SimSetting1}
\bcen
\begin{tabular}{clll} \hline
 Setting & Sample Size & Scaling Factors & {\color{black} Meaning} \\\hline
 1 & $\vn=\vn_1+m\veins_4'$ & $\vsigma=(1,1,1,1)$ & Balanced  homoscedastic \\
 2 & $\vn = \vn_2+m\veins_4'$ & $\vsigma=(1,1,1,1)$ & Unbalanced  homoscedastic \\
 3  &$\vn = \vn_1+m\veins_4'$ & $\vsigma=(1,\sqrt{2},2,\sqrt{5})$ & Balanced  heteroscedastic \\
 4  &$\vn = \vn_2+m\veins_4'$ & $\vsigma=(1,\sqrt{2},2,\sqrt{5})$ &  {\color{black} Unbalanced  heteroscedastic (}Positive Pairing{\color{black})}\\
 5  &$\vn = \vn_2+m\veins_4'$ & $\vsigma=(\sqrt{5},2,\sqrt{2},1)$ & {\color{black} Unbalanced  heteroscedastic (}Negative Pairing{\color{black})}\\\hline \hline
\end{tabular}
\ecen
\end{table}

Since the balanced settings have been discussed in the main paper we here only comment on the unbalanced settings.

In Table~\ref{Table 2} the simulation results for the unbalanced homoscedastic designs (Setting 2) for various distributions are displayed. 
As in Setting~1 the hypothesis $H_0^F$ holds here and it is not surprising that similar observations can be drawn. 
First, the Kruskal-Wallis test controls the nominal type-1 error level ($\alpha = 5\%$) very satisfactorily for all investigated distributions. 
Second, both of the Wald-type statistics (DTS and WTS) tend to be considerably liberal, where their liberality again slowly decreases 
with increasing sample sizes. However, even for the scenarios with larger sample sizes their type-I-error control is not acceptable. 
Finally, the behaviour of the ANOVA-type tests is again similar to the main paper: For smaller sample sizes ($N\leq 120$) both the tests 
$\widehat{\varphi}_N$ and $\widetilde{\varphi}_N$ are slightly liberal. For larger sample sizes their type-$I$ error control is acceptable. 
In contrast, the ANOVA-type test $\varphi_N$ based on the $F$-approximation shows a better control of the type-1 error level and is only slightly liberal 
in case of the smallest simulated sample sizes.

In the two unbalanced heteroscedastic Settings ~4 and 5 the null hypothesis $H_0^F$ is violated and only $H_0^p$ is true. The corresponding results are shown in 
Tables~\ref{Table 4}--\ref{Table 5}.

In case of positive pairings (see the results for Setting 4 in Table~\ref{Table 4}), the Kruskal-Wallis test tends to be 
conservative in all considered scenarios. In comparison to Settings 1 - 3, both the methods 
$\widehat{\varphi}_N$ and $\widetilde{\varphi}_N$ tend to be less liberal and
fairly control the type-1 error rate $\alpha$. The two Wald-type tests (DTS and WTS) are still but less liberal. 
Also in this setup, the ANOVA-type test $\varphi_N$ controls the type-1 error rate very satisfactorily. 

The most severe case from all investigated scenarios is when larger sample sizes
are combined with smaller variances (negative pairing -- Setting 5). The simulation results are 
displayed in Table~\ref{Table 5} below. It can be readily seen that the Kruskal-Wallis test  {\color{black} and both Wald-type tests tend} 
to quite liberal conclusions.  {\color{black}Moreover, b}oth the methods 
$\widehat{\varphi}_N$ and $\widetilde{\varphi}_N$ do not control the error 
rate $\alpha=5\%$  {\color{black}in this set-up.} The method $\varphi_N$
tends to be slightly liberal in case of the smallest simulated sample sizes, but controls the type-1 error rate {\color{black}superior to all other methods.
Thus, the ANOVA-type test $\varphi_N$ is recommended for practical applications.

\text{ }\\[-10ex]
\begin{table}[H]{\color{black}
\caption{Type-I error ($\alpha=5\%$) simulations of the Kruskal-Wallis test (KW), the two Wald-type tests in the test statistics WTS and the test statistic of De Neve and Thas (DTS) and the 
three different ANOVA-type tests $\widehat{\varphi}_N$, $\widetilde{\varphi}_N$, and $\varphi_N$ using the distributional approximations as given in 
 (4.21), (4.23), and (4.24) of the main paper 
under Setting~2 as described in Table~\ref{SimSetting1}. }\label{Table 2}
{\footnotesize\bcen
\begin{tabular}{cccccccc}\hline
Distribution   & Sample Sizes                                     & KW              &  DTS              & WTS              &  $\widehat{\varphi}_N$           &  $\widetilde{\varphi}_N$ & ${\varphi}_N$\\\hline
DExp	&	10	20	30	40	&	0.0518	&	0.1114	&	0.0908	&	0.0758	&	0.0777	&	0.0659	\\
DExp	&	15	25	35	45	&	0.0483	&	0.0926	&	0.0727	&	0.0604	&	0.0607	&	0.0538	\\
DExp	&	20	30	40	50	&	0.0475	&	0.0804	&	0.0740	&	0.0562	&	0.0564	&	0.0515	\\
DExp	&	30	40	50	60	&	0.0497	&	0.0674	&	0.0629	&	0.0559	&	0.0555	&	0.0520	\\
DExp	&	35	45	55	65	&	0.0492	&	0.0650	&	0.0604	&	0.0531	&	0.0538	&	0.0500	\\\hline
LogNor	&	10	20	30	40	&	0.0480	&	0.1112	&	0.0903	&	0.065	&	0.0667	&	0.0580	\\
LogNor	&	15	25	35	45	&	0.0482	&	0.0874	&	0.0730	&	0.0588	&	0.0597	&	0.0526	\\
LogNor	&	20	30	40	50	&	0.0481	&	0.0824	&	0.0711	&	0.0543	&	0.0555	&	0.0498	\\
LogNor	&	30	40	50	60	&	0.0505	&	0.0726	&	0.0654	&	0.0549	&	0.0553	&	0.0510	\\
LogNor	&	35	45	55	65	&	0.0468	&	0.0720	&	0.0696	&	0.0506	&	0.0520	&	0.0476	\\\hline
Normal	&	10	20	30	40	&	0.0477	&	0.1026	&	0.0912	&	0.0650	&	0.0667	&	0.0576	\\
Normal	&	15	25	35	45	&	0.0483	&	0.0916	&	0.0764	&	0.0578	&	0.0588	&	0.0504	\\
Normal	&	20	30	40	50	&	0.0478	&	0.0820	&	0.0708	&	0.0517	&	0.0515	&	0.0475	\\
Normal	&	30	40	50	60	&	0.0518	&	0.0732	&	0.0645	&	0.0559	&	0.0570	&	0.0525	\\
Normal	&	35	45	55	65	&	0.0531	&	0.0712	&	0.0639	&	0.0579	&	0.0583	&	0.0535	\\\hline
\end{tabular}
\ecen}
}
\end{table}

\begin{table}[H]{\color{black}
\caption{Type-I error ($\alpha=5\%$) simulations of the Kruskal-Wallis test (KW), the two Wald-type tests in the test statistics WTS and the test statistic of De Neve and Thas (DTS) and the 
three different ANOVA-type tests $\widehat{\varphi}_N$, $\widetilde{\varphi}_N$, and $\varphi_N$ using the distributional approximations as given in 
(4.21), (4.23), and (4.24) of the main paper under Setting~4 as described in Table~\ref{SimSetting1}. }\label{Table 4}
{\footnotesize\bcen
\begin{tabular}{cccccccc}\hline
Distribution   & Sample Sizes                                     & KW              &  DTS              & WTS              &  $\widehat{\varphi}_N$           &  $\widetilde{\varphi}_N$ & ${\varphi}_N$\\\hline
DExp	&	10	20	30	40	&	0.0247	&	0.1008	&	0.0749	&	0.0563	&	0.0566	&	0.0492	\\
DExp	&	15	25	35	45	&	0.0318	&	0.0822	&	0.0696	&	0.0534	&	0.0533	&	0.0486	\\
DExp	&	20	30	40	50	&	0.0371	&	0.0828	&	0.0682	&	0.0578	&	0.0573	&	0.0541	\\
DExp	&	30	40	50	60	&	0.0414	&	0.0730	&	0.0640	&	0.0543	&	0.0549	&	0.0515	\\
DExp	&	35	45	55	65	&	0.0421	&	0.0696	&	0.0584	&	0.0525	&	0.0533	&	0.0498	\\\hline
LogNor	&	10	20	30	40	&	0.0364	&	0.1000	&	0.0873	&	0.0653	&	0.0670	&	0.0586	\\
LogNor	&	15	25	35	45	&	0.0392	&	0.0904	&	0.0736	&	0.0567	&	0.0562	&	0.0511	\\
LogNor	&	20	30	40	50	&	0.0387	&	0.0752	&	0.0686	&	0.0538	&	0.0546	&	0.0495	\\
LogNor	&	30	40	50	60	&	0.0420	&	0.0718	&	0.0665	&	0.0527	&	0.0535	&	0.0494	\\
LogNor	&	35	45	55	65	&	0.0407	&	0.0684	&	0.0597	&	0.0507	&	0.0511	&	0.0479	\\\hline
Normal	&	10	20	30	40	&	0.0248	&	0.0938	&	0.0706	&	0.0545	&	0.0547	&	0.0475	\\
Normal	&	15	25	35	45	&	0.0287	&	0.0858	&	0.0710	&	0.0525	&	0.0524	&	0.0470	\\
Normal	&	20	30	40	50	&	0.0340	&	0.0834	&	0.0674	&	0.0546	&	0.0549	&	0.0493	\\
Normal	&	30	40	50	60	&	0.0411	&	0.0690	&	0.0652	&	0.0523	&	0.0528	&	0.0492	\\
Normal	&	35	45	55	65	&	0.0432	&	0.0702	&	0.0587  &	0.0512	&	0.0517	&	0.0479	\\\hline
				\end{tabular}
				\ecen}}
\end{table}

\begin{table}[H]{\color{black}
\caption{Type-I error ($\alpha=5\%$) simulations of the Kruskal-Wallis test (KW), the two Wald-type tests in the test statistics WTS and the test statistic of De Neve and Thas (DTS) and the 
three different ANOVA-type tests $\widehat{\varphi}_N$, $\widetilde{\varphi}_N$, and $\varphi_N$ using the distributional approximations as given in 
(4.21), (4.23), and (4.24) of the main paper under Setting~5 as described in Table~\ref{SimSetting1}. }\label{Table 5}
{\footnotesize\bcen
\begin{tabular}{cccccccc}\hline
Distribution   & Sample Sizes                                     & KW              &  DTS              & WTS              &  $\widehat{\varphi}_N$           &  $\widetilde{\varphi}_N$ & ${\varphi}_N$\\\hline
DExp	&	10	20	30	40	&	0.1178	&	0.1108	&	0.0907	&	0.0755	&	0.0760	&	0.0628	\\
DExp	&	15	25	35	45	&	0.1029	&	0.1008	&	0.0768	&	0.0652	&	0.0665	&	0.0580	\\
DExp	&	20	30	40	50	&	0.1014	&	0.0916	&	0.0721	&	0.0602	&	0.0609	&	0.0538	\\
DExp	&	30	40	50	60	&	0.0888	&	0.0722	&	0.0671	&	0.0544	&	0.0554	&	0.0502	\\
DExp	&	35	45	55	65	&	0.0879	&	0.0746	&	0.0655	&	0.0544	&	0.0557	&	0.0496	\\\hline
LogNor	&	10	20	30	40	&	0.0695	&	0.1060	  &	0.0912	&	0.0753	&	0.0760	&	0.0644	\\
LogNor	&	15	25	35	45	&	0.0667	&	0.0962	&	0.0795	&	0.0654	&	0.0660	&	0.0583	\\
LogNor	&	20	30	40	50	&	0.0580	&	0.0796	&	0.0685	&	0.0550	&	0.0557	&	0.0508	\\
LogNor	&	30	40	50	60	&	0.0586	&	0.0768	&	0.0629	&	0.0497	&	0.0506	&	0.0465	\\
LogNor	&	35	45	55	65	&	0.0623	&	0.0666	&	0.0638	&	0.0563	&	0.0565	&	0.0515	\\\hline
Normal	&	10	20	30	40	&	0.1287	&	0.1132	&	0.0935	&	0.0719	&	0.0727	&	0.0619	\\
Normal	&	15	25	35	45	&	0.1198	&	0.0882	&	0.0798	&	0.0656	&	0.0675	&	0.0583	\\
Normal	&	20	30	40	50	&	0.1167	&	0.0800	&	0.0713	&	0.0624	&	0.0635	&	0.0549	\\
Normal	&	30	40	50	60	&	0.1044	&	0.0724	&	0.0678	&	0.0598	&	0.0608	&	0.0546	\\
Normal	&	35	45	55	65	&	0.0992	&	0.0746	&	0.0617	&	0.0562	&	0.0571	&	0.0510	\\\hline
\end{tabular}
     \ecen}}
\end{table}

\newpage

\bcen
{\sc \large REFERENCES}
\ecen

\bdes\itemsep.5pt

{\footnotesize

\item {\sc Akritas, M.~G., and \ Arnold, S.~F.} (1994). Fully
nonparametric hypotheses for factorial designs I:  Multivariate repeated
measures designs. {\em J.\  Amer.\  Statist.\  Assoc.} {\bf  89},
336--343.

\item {\sc Akritas, M.~G., Arnold, S.~F., and Brunner, E.} (1997).
Nonparametric hypotheses and rank statistics for unbalanced factorial
designs. {\em J.\  Amer.\ Statist.\ Assoc.} {\bf 92}, 258--265.

\item {\sc Akritas, M.~G. and  Brunner, E.} (1997). A unified approach to ranks tests in mixed models. {\em J.\  Statist.\ Plann.\ Inference} {\bf 61}, 249--277.

\item {\sc Akritas, M.~G.} (2011). Nonparametric Models for ANOVA and ANCOVA 
Designs. In {\em International Encyclopedia of Statistical Science}, Springer, 
964--968.

\item {\sc Brunner, E., Munzel, U. and Puri, M.~L.} (1999). Rank-Score Tests in Factorial Designs with Repeated Measures. {\em J. Mult. Analysis} {\bf 70}, 286--317. 

\item {\sc Brunner, E., and Puri, M.~L.} (2001). Nonparametric methods in 
factorial designs. {\em Statistical Papers} {\bf 42}, 1--52.

\item {\sc Brunner, E., Konietschke, F., Pauly, M. and Puri, M.L.} (2016). 
Rank-Based Procedures in Factorial Designs: Hypotheses about Nonparametric Treatment Effects.

\item {\sc De Neve, J., and Thas, O.} (2015). A Regression Framework for Rank 
Tests Based on the Probabilistic Index Model. {\em J.\  Amer.\ Statist.\  
Assoc.}, DOI: 10.1080/01621459.2015.101622.

\item {\sc Domhof, S.} (2001). Nichtparametrische relative Effekte. Ph.D. 
Thesis, University of G\"ottingen. 

\item {\sc Gao, X., and Alvo, M.} (2005). 
A unified nonparametric approach for unbalanced factorial designs. {\em J. 
Amer. Statist. Assoc.} {\bf 100}, 926--941.

\item {\sc Gao, X., and Alvo, M.} (2008). 
Nonparametric multiple comparison procedures for unbalanced two-way layouts.
{\em Journal of Statistical Planning and Inference} {\bf 138}, 3674--3686.

\item {\sc Gao, X., Alvo, M., Chen, J., and Li, G.} (2008). 
Nonparametric multiple comparison procedures for unbalanced one-way factorial
designs. {\em Journal of Statistical Planning and Inference} {\bf 138}, 
2574--2591.

\item {\sc Konietschke, F., Bathke, A. C., Hothorn, L. A., and Brunner, E.} (2010). 
Testing and estimation of purely nonparametric effects in repeated measures designs. 
{\em Computational Statistics and Data Analysis}, {\bf 54}, 1895--1905.

\item {\sc Placzek, M.} (2013). Nichtparametrische simultane Inferenz f\"ur faktorielle Repeated Measures Designs. Master Thesis, 
University of G\"ottingen. 
}
\edes

\end{document}